\documentclass[a4paper, 11pt]{article}\usepackage[]{graphicx}\usepackage[]{color}
\makeatletter
\def\maxwidth{ %
  \ifdim\Gin@nat@width>\linewidth
    \linewidth
  \else
    \Gin@nat@width
  \fi
}
\makeatother

\definecolor{fgcolor}{rgb}{0.345, 0.345, 0.345}

\usepackage{framed}
\makeatletter
 {\par\unskip\endMakeFramed%
 \at@end@of@kframe}
\makeatother

\definecolor{shadecolor}{rgb}{.97, .97, .97}
\definecolor{messagecolor}{rgb}{0, 0, 0}
\definecolor{warningcolor}{rgb}{1, 0, 1}
\definecolor{errorcolor}{rgb}{1, 0, 0}
\newenvironment{knitrout}{}{} % an empty environment to be redefined in TeX

\usepackage{alltt}
\usepackage[T1]{fontenc}
\usepackage[utf8]{inputenc}
\usepackage[english]{babel}
\usepackage{graphics}
\usepackage[dvipsnames]{xcolor}
\usepackage{amsmath, amssymb}
\usepackage{textcomp}
\usepackage[sc]{mathpazo}
\usepackage{helvet}
\usepackage[color=yellow]{todonotes}
\usepackage{doi}
\usepackage[round]{natbib}
\usepackage{xifthen}            % for \isempty
\usepackage{array}
\usepackage{amssymb}

\newcommand{\blanco}[1]{  } 
\newcommand{\latin}[1]{\textit{#1}}

\newcommand{\abk}[1]{\mbox{#1}\xdot}
\DeclareRobustCommand\xdot{\futurelet\token\Xdot}
\def\Xdot{%
  \ifx\token\bgroup.%
  \else\ifx\token\egroup.%
  \else\ifx\token\/.%
  \else\ifx\token\ .%
  \else\ifx\token!.%
  \else\ifx\token,.%
  \else\ifx\token:.%
  \else\ifx\token;.%
  \else\ifx\token?.%
  \else\ifx\token/.%
  \else\ifx\token'.%
  \else\ifx\token).%
  \else\ifx\token-.%
  \else\ifx\token+.%
  \else\ifx\token~.%
  \else\ifx\token.%
  \else.\ %
  \fi\fi\fi\fi\fi\fi\fi\fi\fi\fi\fi\fi\fi\fi\fi\fi%
}

\newcommand{\eg}{\abk{\latin{e.\,g}}}
\newcommand{\ie}{\abk{\latin{i.\,e}}}

\newlength{\halbebreite}
\DeclareMathOperator{\Bin}{Bin} % Binomial Distribution - 
\DeclareMathOperator{\Nor}{N} % Normal -
\DeclareMathOperator{\Be}{Be} % Beta - 
\DeclareMathOperator{\Var}{Var} % Varianz
\DeclareMathOperator{\sign}{sign} % signum
\renewcommand{\P}{\operatorname{\mathsf{Pr}}} % Wahrscheinlichkeitsmaß
\DeclareMathOperator{\BF}{BF} % Bayes factor
\newcommand{\partialv}[3][1]{%
\ifthenelse{#1 = 1}{\frac{\partial #2}{\partial #3}}{\frac{\partial^{#1} #2}{\partial #3^{#1}}}
} 

\newcommand{\partials}[3][1]{%
\ifthenelse{#1 = 1}{\frac{d #2}{d #3}}{\frac{d^{#1} #2}{d #3^{#1}}}
} 

\newcommand{\dseps}[2][1]{%
\ifthenelse{#1 = 1}{\frac{d}{d #2}}{\frac{d^{#1}}{d #2^{#1}}}
}

\newcommand{\dsepv}[2][1]{%
\ifthenelse{#1 = 1}{\frac{\partial}{\partial #2}}{\frac{\partial^{#1}}{\partial #2^{#1}}}
}

\newcommand{\ml}[2][1]{% % für Maximum-Likelihood-Schätzer von #1
\ifthenelse{#1 = 1}%
 {\hat{#2}_{\scriptscriptstyle{\mathrm{ML}}}}% 
 {\hat{#2}^{#1}_{\scriptscriptstyle{\mathrm{ML}}}}% z.B. für sigmadach^2
}
\newcommand{\map}[2][0]{% % für MAP-Schätzer von #1
\ifthenelse{#1 = 0}%
 {\hat{#2}_{\scriptscriptstyle{\mathrm{MAP}}}}% 
 {\hat{#2}_{{\scriptscriptstyle{\mathrm{MAP}}_{#1}}}}% z.B. für sigmadach^2
}
\newcommand{\mpm}[2][0]{% % für MPM-Schätzer von #1
\ifthenelse{#1 = 0}%
 {\hat{#2}_{\scriptscriptstyle{\mathrm{MPM}}}}% 
 {\hat{#2}_{{\scriptscriptstyle{\mathrm{MPM}}_{#1}}}}% z.B. für sigmadach^2
}

\newcommand{\given}{\,\vert\,} % für "X gegeben Y" also $X\given Y$ schreiben
\newcommand{\abs}[1]{\left\lvert#1\right\rvert} % Absolutbetrag
\usepackage{booktabs} % nicer tables
\usepackage{pdflscape} % allow landscape mode for pdfs
\usepackage[toc, page]{appendix}
\usepackage{nameref}
\usepackage{enumitem}
\usepackage{caption}
 
\newcommand{\that}{\hat{\theta}}
\newcommand{\ps}{p_{\text{S}}}
\newcommand{\BFs}{\BF_{\text{S}}}
\newcommand{\BFr}{\BF_{\text{R}}}

\newcommand{\lw}[1]{W_{\scriptscriptstyle{#1}}}

\newcommand\BFcol[2]{\BF_{\scriptscriptstyle{#1:#2}}}
\newcommand{\gy}{g_{\scriptstyle \gamma}}
\newcommand{\dmin}{d_{\text{min}}}
\newcommand{\dmax}{d_{\text{max}}}

\newcommand{\HS}{H_{\text{S}}}
\newcommand{\HA}{H_{\text{A}}}
\newcommand{\HAp}{H_{\text{A}^\prime}}

\newcommand{\minBF}{\text{minBF}}
\newcommand{\gminBFo}{g_{\scriptscriptstyle \minBF_{\scriptscriptstyle o}}}

\usepackage{geometry}
\title{\vspace{-1em}\textbf{The sceptical Bayes factor for the
assessment of replication success}}
\author{\textbf{Samuel Pawel, Leonhard Held} \\
Epidemiology, Biostatistics and Prevention Institute (EBPI) \\
Center for Reproducible Science (CRS) \\
University of Zurich, Switzerland \\
E-mail: \href{mailto:samuel.pawel@uzh.ch}{samuel.pawel@uzh.ch}}
\date{August 23, 2021} % hard-code date for arXiv submission

\usepackage{hyperref}
\hypersetup{
  bookmarksopen=true,
  breaklinks=true,
  pdftitle={The sceptical Bayes factor},
  pdfauthor={Samuel Pawel},
  pdfsubject={},
  pdfkeywords={},
  colorlinks=true,
  linkcolor=RoyalPurple,
  anchorcolor=black,
  citecolor=MidnightBlue,
  urlcolor=BrickRed,
}

\usepackage{fancyhdr}
\IfFileExists{upquote.sty}{\usepackage{upquote}}{}
\begin{document}
\maketitle

% Abstract
\begin{center}
\begin{minipage}{14cm}
{\footnotesize \rule{\textwidth}{0.5pt} \\ {\centering \textbf{Abstract}
    \\ Replication studies are increasingly conducted but there is no
established statistical criterion for replication success. We propose a novel
approach combining reverse-Bayes analysis with Bayesian hypothesis testing: a
sceptical prior is determined for the effect size such that the original finding
is no longer convincing in terms of a Bayes factor. This prior is then
contrasted to an advocacy prior (the reference posterior of the effect size
based on the original study), and replication success is declared if the
replication data favour the advocacy over the sceptical prior at a higher level
than the original data favoured the sceptical prior over the null hypothesis.
The sceptical Bayes factor is the highest level where replication success can be
declared. A comparison to existing methods reveals that the sceptical Bayes
factor combines several notions of replicability: it ensures that both studies
show sufficient evidence against the null and penalises incompatibility of their
effect estimates. Analysis of asymptotic properties and error rates, as well as
case studies from the Social Sciences Replication Project show the advantages of
the method for the assessment of replicability.
}
\rule{\textwidth}{0.4pt} 
}
\end{minipage}
\end{center}

% Introduction
\section{Introduction}
As a consequence of the so-called replication crisis, the scientific community
increasingly recognises the value of replication studies, and several attempts
have been made to assess replicability on a large scale \citep{Errington2014,
Klein2014, Opensc2015, Camerer2016, Camerer2018, Cova2018}. Despite most
researchers agreeing on the importance of replication, there is currently no
agreement on a statistical criterion for replication success. Instead, a variety
of statistical methods, frequentist \citep{Simonsohn2015, Patil2016, Hedges2019,
Mathur2020}, Bayesian \citep{Bayarri2002b, Bayarri2002, Verhagen2014,
Johnson2016, Etz2016, vanAert2017, Ly2018, Harms2019}, and combinations thereof
\citep{Held2020, Pawel2020, Held2021} have been proposed to quantify replication
success.

Due to this lack of an established method, replication projects typically report
the results of several methods and it is not uncommon for these to contradict
each other. For example, both studies may find evidence against a null effect,
but the individual effect estimates may still be incompatible (often the
replication estimate is much smaller). Conversely, both estimates may be
compatible, but there may not be enough evidence against a null effect in one of
the studies.

The objective of this paper is to present a novel Bayesian method for
quantifying replication success, which builds upon a previously proposed method
\citep[the \emph{sceptical $p$-value} from][]{Held2020} and unifies several
notions of replicability. The method combines the natural fit of the
reverse-Bayes approach to the replication setting with the use of Bayes factors
for hypothesis testing \citep{Jeffreys1961, Kass1995} and model criticism
\citep{Box1980}. In a nutshell, replication success is declared if the
replication data favour an advocacy prior for the effect size, which emerges
from taking the original result at face value, over a sceptical prior, which
renders the original result unconvincing.

\citet{Held2020} proposed a reverse-Bayes approach for the assessment of
replication success: The main idea is to challenge the result from an original
study by determining a \emph{sceptical prior} for the effect size, sufficiently
concentrated around the null value such that the resulting posterior is rendered
unconvincing \citep{Matthews2001b}. An unconvincing posterior at level $\alpha$
is defined by its $(1 - \alpha)$ credible interval just including the null
value. Subsequently, the replication data are used in a prior-data conflict
assessment \citep{Box1980, Evans2006} and replication success is concluded if
there is sufficient conflict between the sceptical prior and the replication
data. Specifically, replication success at level $\alpha$ is established if the
prior predictive tail probability of the replication estimate is smaller than
$\alpha$. The smallest level $\alpha$ at which replication success can be
declared corresponds to the sceptical $p$-value.

The method comes with appealing properties: The sceptical $p$-value is never
smaller than the ordinary $p$-values from both studies, thus ensuring that they
both provide evidence against the null. At the same time, it also takes into
account the size of their effect estimates, penalising the case when the
replication estimate is smaller than the original estimate. \citet{Held2021}
further refined the method with a recalibration that allows the sceptical
$p$-value to be interpreted on the same scale as an ordinary $p$-value, as well
as ensuring appropriate frequentist properties, such as type-I error rate
control if the replication sample size is not smaller than in the original
study.

Despite the methods' Bayesian nature, it relies on tail probabilities as primary
inference tool. An alternative is the Bayes factor, the principled Bayesian
solution to hypothesis testing and model selection \citep{Jeffreys1961,
Kass1995}. In contrast to tail probabilities, Bayes factors have a more natural
interpretation and allow for direct quantification of evidence for one
hypothesis versus another. In this paper we therefore extend the reverse-Bayes
procedure from \citet{Held2020} to use Bayes factors for the purpose of
quantifying evidence. This extension was suggested by \citet{Consonni2019} and
\citet{Pericchi2020} independently. Interestingly, a similar extension of the
reverse-Bayes method from \citet{Matthews2001b} was already hinted at by
\citet{Berger2001}, but to date no one has attempted to realise the idea.

The inclusion of Bayes factors leads to a new quantity which we call the
\emph{sceptical Bayes factor}. Unlike standard forward-Bayes methods, but
similar to the sceptical $p$-value, the proposed method combines two notions of
replication success: It requires from both studies to show sufficient evidence
against the null, while also penalising incompatibility of their effect
estimates. However, while the sceptical $p$-value quantifies compatibility only
indirectly through conflict with the sceptical prior, the sceptical Bayes factor
evaluates directly how likely the replication data are to occur under an
advocacy prior (the reference posterior of the effect conditional on the
original study). This direct assessment of compatibility allows for stronger
statements about the degree of replication success, and it may also lead to
different conclusions in certain situations.

This paper is structured as follows: Section \ref{sec:methods} presents the
derivation of the sceptical Bayes factor. Its asymptotic and finite sample
properties are then compared with other measures of replication success in
Section \ref{sec:comparison}. An extension to non-normal models is presented in
Section \ref{sec:tdist}. Section \ref{sec:applications} illustrates how the
method works in practice using case studies from % the Reproducibility Project:
Cancer Biology % \citep{Errington2014} and the Social Sciences Replication
Project \citep{Camerer2018}. Section \ref{sec:discussion} provides concluding
remarks about strengths, limitations and extensions of the method.

\subsection*{Notation and assumptions}
Denote the Bayes factor comparing the plausibility of hypotheses $H_1$ and $H_2$
with respect to the observed data $x$ by
\begin{align*}
  \BFcol{1}{2}(x) = \frac{f(x \given H_1)}{f(x \given H_2)}
  % = \frac{\P(H_1 \given x)}{\P(H_2 \given x)} \cdot \frac{\P(H_2)}{\P(H_1)}
  = \frac{\int_{\Theta_1} f(x \given \theta_1)f(\theta_1) \, \text{d}\theta_1}{
  \int_{\Theta_2} f(x \given \theta_2)f(\theta_2) \, \text{d}\theta_2},
\end{align*}
where $f(x \given H_i)$ is the marginal likelihood of the data under $H_i$
obtained by integrating the likelihood $f(x \given \theta_{i})$ with respect to
the prior distribution $f(\theta_{i})$ of the model parameters
$\theta_i \in \Theta_i$ with $i = 1, 2$. Sometimes we will also write
$\BFcol{1}{2}(x ; \phi^\prime)$ to indicate that the Bayes factor is evaluated
for a specific value $\phi^\prime$ of a hyperparameter $\phi$ of one of the
model priors. To simplify comparison with $p$-values we will orient Bayes
factors such that lower values indicate more evidence against a null hypothesis.

Let $\theta$ denote the effect of a treatment on an outcome of interest. Let
$\that_o$ and $\that_r$ denote its maximum likelihood estimates obtained from an
original (subscript $o$) and from a replication study (subscript $r$),
respectively. Let the corresponding standard errors be denoted by $\sigma_o$ and
$\sigma_r$, the $z$-values by $z_o = \that_o/\sigma_o$ and
$z_r = \that_r/\sigma_r$, and define the variance ratio as
$c = \sigma_o^2/\sigma_r^2$ and the relative effect estimate as
$d = \that_r/\that_o = z_r/(z_o\sqrt{c})$. For many effect size types the
variances are inversely proportional to the sample size, \ie
$\sigma^2_{o} = \kappa/n_{o}$ and $\sigma^2_{r} = \kappa/n_{r}$ for some unit
variance $\kappa$. The variance ratio is then the ratio of the replication to
the original sample size $c = n_{r}/n_{o}$.

We adopt a meta-analytic framework and consider the effect estimates as the
data, rather than their underlying samples, and assume that
$\that_k \given \theta \sim \Nor(\theta, \sigma^2_k)$ for $k \in \{o, r\},$ \ie
normality of the effect estimates around $\theta$, with known variances equal to
their squared standard errors. For studies with reasonable sample size, this
framework usually provides a good approximation for a wide range of (suitably
transformed) effect size types \citep[Chapter 2.4]{Spiegelhalter2004}. For
example, means and mean differences (no transformation), odds ratios, hazard
ratios, risk ratios (logarithmic transformation), or correlation coefficients
(``Fisher-$z$'' transformation). We refer to the literature of meta-analysis for
details about transformations of effect sizes \citep[\eg][Chapter
11.6]{Cooper2019}. The normal model in combination with conjugate priors enables
derivation of closed-form expressions in many cases, which allows us to easily
study limiting behaviour and facilitates interpretability. In Section
\ref{sec:tdist}, we will present relaxations of the normality assumption, which
can lead to more accurate inferences when studies have small sample sizes and/or
show extreme results.

\section{Reverse-Bayes assessment of replication success with Bayes factors}
\label{sec:methods}
\begin{figure}[!htb]
  \centering
%% first attempt
\begin{tikzpicture}[node distance = 5em, thick]
  % nodes %
  \node[font = \large] (data) {$f(\text{data} \,|\, \theta)$};
  \node[font = \large] (prior) [left=7em of data] {$f(\theta)$};
  \node[font = \large] (post) [right=7em of data] {$f(\theta \,|\, \text{data})$};
  \node[text width=10em,text centered] (RBinference) [below of=prior] {Reverse-Bayes inference};
  \node[text width=10em, text centered] (FBinference) [below of=post] {Forward-Bayes inference};

  % edges %
  \draw [->, dotted] (prior) to [out=45, in=135, looseness=0.3] node[above]{Bayesian updating} (post);
  %% other names could be:
  %% - reverse Bayesian updating vs forward Bayesian updating
  %% - Bayesian reverting
  \draw [->, dashed] (post) to [out=225, in=315, looseness=0.3] node[below]{Bayesian downdating} (prior);
  \draw [->, dashed] (prior) to [out=270, in=90, looseness=0] (RBinference);
  \draw [->, dotted] (post) to [out=270, in=90, looseness=0] (FBinference);
\end{tikzpicture}
\caption{Schematic illustration of reverse-Bayes and forward-Bayes inference.}
\label{fig:RBschmema}
\end{figure}
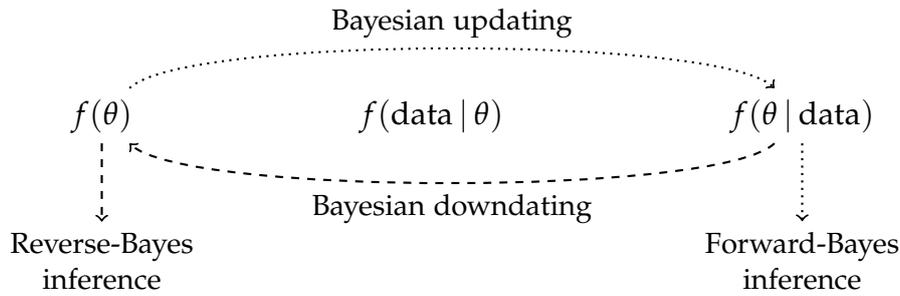

% ``judgements are made concerning the values that the final
% probabilities would have after some imaginary results are obtained; then Bayes'
% theorem is used backwards to estimate what the consistent initial probabilities
% must be assumed to be; and lastly Bayes's theorem is used (forwards) to obtain
% estimates of the final probabilities in the light of the actual observations.
% This device takes \emph{seriously} the view that a theory of probability or of
% rationality is no more than a theory of consistency'' (p. 126).
The idea of reversing Bayes' theorem was first proposed by \citet{Good1950}. He
acknowledged that in many situations there is no obvious choice for the prior
distributions involved in Bayesian analyses. On the other hand, we are often
more certain which posterior inferences would convince us regarding the
credibility of a hypothesis. For this reason, Good inverted Bayes' theorem and
derived priors, which combined with the observed data, would lead to posterior
inferences that were specified beforehand (\eg the data favour one hypothesis
over another). His reverse-Bayes inference then centred around the question
whether the resulting prior is plausible, and if so, this would legitimise the
posterior inference. See Figure~\ref{fig:RBschmema} for a graphical illustration
of this process.

Good argued that philosophically there is nothing wrong with inferences
resulting from backwards use of Bayes' theorem, since the theorem merely
constrains prior and posterior to be consistent with the laws of probability
(regardless of their conventional names suggesting a particular temporal
ordering). Despite his advocacy, the reverse-Bayes idea remained largely
unexplored until \citet{Matthews2001b} introduced the \emph{Analysis of
Credibility}, which in turn led to new developments in reverse-Bayes methodology
(see \citet{Held2021b} for a recent review). Most of these approaches use the
reversal of Bayes' theorem in order to challenge or substantiate the credibility
of scientific claims. Usually, a posterior inference corresponding to
(non-)credibility of a claim is specified, and the associated prior is then
derived from the data. Inference is subsequently carried out based on this
reverse-Bayes prior, \eg the interest is often to check whether the prior is
plausible in light of external evidence, an obvious candidate being data from a
replication study. This can be done, for example, using methods to assess
prior-data conflict \citep{Box1980, Evans2006}.

% Good, Good Thinking, P. 126:
%``In some cases it should be possible to make some progress in the estimation of
% relative initial probabilities by assuming that the current judgements are correct and
% by using Bayes's theorem backwards. This idea is closely related to the ``device
% of imaginary results'' in which judgements are made concerning the values that the
% final probabilities would have after some imaginary results are obtained; then Bayes'
% theorem is used backwards to estimate what the consistent initial probabilities must
% be assumed to be; and lastly Bayes's theorem is used (forwards) to obtain estimates
% of the final probabilities in the light of the actual observations. This device takes
% \emph{seriously} the view that a theory of probability or of rationality is no more
% than a theory of consistency.''

In this paper, we consider a reverse-Bayes procedure consisting of two stages
that naturally fit the replication setting: We first determine a
\emph{sufficiently sceptical prior} for the effect $\theta$ such that the
original result is no longer convincing in terms of a suitable Bayes factor.
Using another Bayes factor, we then quantify replication success by comparing
how likely the replication data are predicted by the sufficiently sceptical
prior relative to an \emph{advocacy prior}, which is the posterior of the effect
$\theta$ conditional on the original data and an uninformative/reference prior.
Box \hyperref[box:nutshell]{1} provides a summary of the procedure, the
following sections will explain it in more detail.

% summarize procedure in box
\begin{table}[!hbt]
\begin{center}
\fbox{
\begin{minipage}{.85\linewidth}
  \begin{enumerate}[leftmargin=*]

    \item \textbf{Original study}:
    For the original effect estimate
$\that_o \given \theta \sim \Nor(\theta, \sigma^2_o)$ consider the point null
hypothesis $H_0 \colon \theta = 0$ vs. $\HS \colon \theta \neq 0$. Fix a level
$\gamma \in (0, 1)$ and determine the sufficiently sceptical prior under the
alternative $\theta \given \HS \sim \Nor(0, \gy \cdot \sigma^2_o)$ such that the
Bayes factor contrasting $H_0$ to $\HS$
is $$\BFcol{0}{\text{S}}(\hat{\theta}_o; \gy) = \gamma.$$ The prior
$\theta \given \HS$ represents a \emph{sceptic} who remains unconvinced about
the presence of an effect at level $\gamma$.

    \item \textbf{Replication study}:
          For the replication effect estimate
$\that_r \given \theta \sim \Nor(\theta, \sigma^2_r)$ compute the Bayes factor
contrasting the sceptic $\HS \colon \theta \sim \Nor(0, \gy \cdot \sigma^2_o)$
to an advocate $\HA \colon \theta \sim \Nor(\that_o, \sigma^2_o)$. Declare
\emph{replication success} at level $\gamma$ if
          $$\BFcol{\text{S}}{\text{A}}(\that_r; \gy) \leq \gamma,$$
          \ie the data favour the advocate over the sceptic at a higher level
than the sceptic's initial objection.

    \item[$\rightarrow$] The \emph{sceptical Bayes factor} $\BFs$ is the
smallest level $\gamma$ at which replication success can be declared.
 \end{enumerate}
\end{minipage}
}
\end{center}
\caption*{Box 1: Summary of reverse-Bayes assessment of replication success with
Bayes factors.}
\label{box:nutshell}
\end{table}

\subsection{Data from the original study}
For the effect estimate $\that_o \given \theta \sim \Nor(\theta, \sigma^2_o)$
from the original study consider a hypothesis test comparing the null hypothesis
$H_0\colon$ $\theta = 0$ to the alternative $\HS \colon$ $\theta \neq 0$.
Specification of a prior distribution for $\theta$ under $\HS$ is now required
for Bayesian hypothesis testing. A typical choice \citep{Jeffreys1961} is a
local alternative, a unimodal symmetric prior distribution centred around the
null value. We consider the sceptical prior
$\theta \given \HS \sim \Nor(0, \sigma^2_s = g \cdot \sigma^2_o)$ with relative
sceptical prior variance $g$ for this purpose (relative to the variance from the
original estimate $g = \sigma^2_s/\sigma^2_o$), resembling the $g$-prior known
from the regression literature \citep{Zellner1986, Liang2008}. The explicit form
of the Bayes factor is then given by
\begin{align}
  \label{eq:BFo}
  \BFcol{0}{\text{S}}(\hat{\theta}_o; g)
  = \sqrt{1 + g} \cdot
  \exp\left\{-\frac{1}{2} \cdot \frac{g}{1 + g} \cdot z^2_o\right\}.
\end{align}

\begin{figure}[!htb]
\begin{knitrout}
\definecolor{shadecolor}{rgb}{0.969, 0.969, 0.969}\color{fgcolor}
\includegraphics[width=\maxwidth]{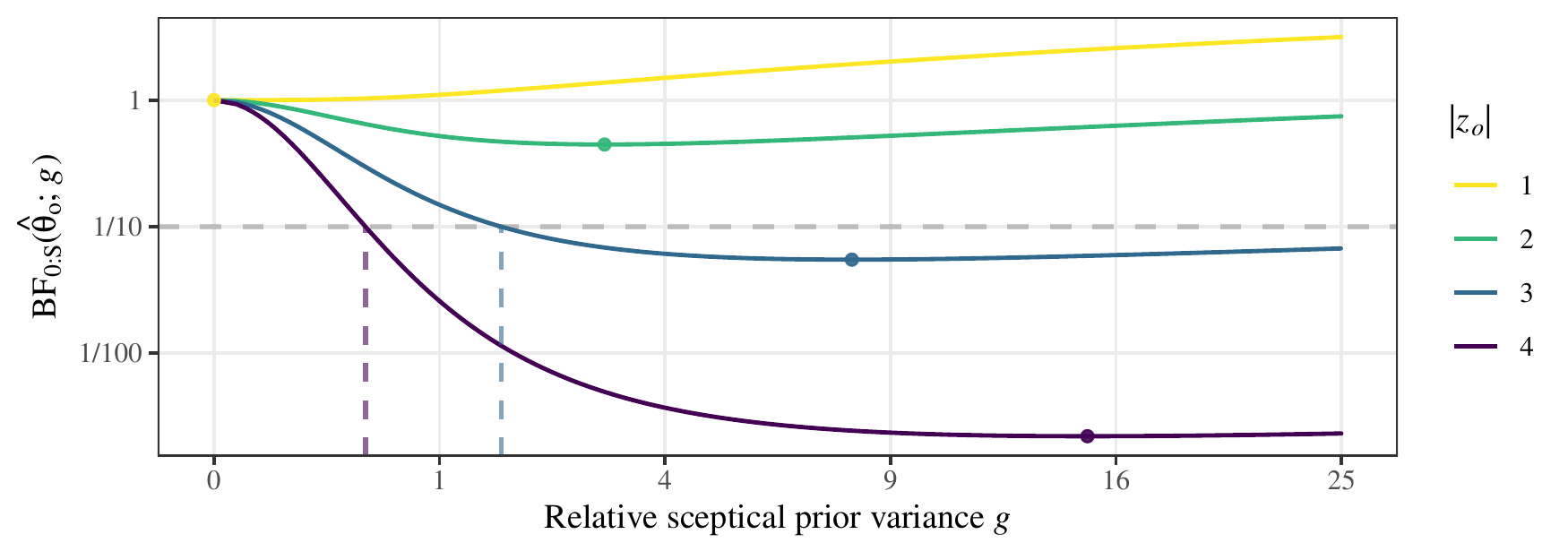}
\end{knitrout}
\caption{Bayes factor $\BFcol{0}{\text{S}}(\hat{\theta}_o; g)$ as a function of
relative sceptical prior variance $g$ for different values of
$|z_o| = |\that_o|/\sigma_o$. Minimum Bayes factors $\minBF_{o}$ are indicated
by dots. Dashed vertical lines indicate sufficiently sceptical relative prior
variance $g_{\scriptscriptstyle \gamma}$ at level $\gamma = 1/10$, if they
exist.}
\label{fig:bfo}
\end{figure}

The Bayes factor from equation ~\eqref{eq:BFo} is shown in Figure ~\ref{fig:bfo}
as a function of $g$ and for different original $z$-values $z_o$. For fixed
$z_o$, it is well known that this Bayes factor is bounded from below by
\begin{align}
  \label{eq:minBFo}
  \minBF_{o} =
  \begin{cases}
    |z_o| \cdot  \exp(-z_o^2/2) \cdot \sqrt{e} & \text{for} ~ |z_o| > 1 \\
    1 & \text{for} ~ |z_o| \leq 1
  \end{cases}
\end{align}
which is reached at $\gminBFo = \max\{0, z^2_o - 1\}$ \citep{Edwards1963}.
Further increasing the relative sceptical prior variance increases
\eqref{eq:BFo} indefinitely because of the Jeffreys-Lindley paradox, \ie
$\BFcol{0}{\text{S}}(\hat{\theta}_o; g) \to \infty$ for $g \to \infty$
\citep[][Section 6.1.4]{Bernardo2000}. Hence, for a relative sceptical prior
variance $g \in [0, \gminBFo]$, the resulting Bayes factor will be
$\BFcol{0}{\text{S}}(\hat{\theta}_o; g) \in [\minBF_o, 1]$.

We now apply the reverse-Bayes idea and challenge the original finding. To do
so, we fix a level $\gamma$ above which the original finding is no longer
convincing to us. For example, this could be $\gamma = 1/10$; the threshold for
strong evidence against $H_0$ according to the classification from
\citet{Jeffreys1961}. Suppose now there exists a $\gy \leq \gminBFo$ such that
$\BFcol{0}{\text{S}}(\hat{\theta}_o; \gy) = \gamma$. It can be shown (Appendix
\ref{appendix:ssrv}) that $\gy$ can be explicitly computed by
\begin{align}
\label{ggamma}
  \gy &=
  \begin{cases}
    -\dfrac{z_o^2}{q} - 1 & ~~ \text{if} ~ -\dfrac{z_o^2}{q} \geq 1 \\
    \text{undefined} & ~~ \text{else}
  \end{cases} \\
  \text{where} ~ q &= \lw{-1} \left(-\frac{z_o^2}{\gamma^2} \cdot
  \exp\left\{-z_o^2\right\}\right)
  \nonumber
\end{align}
with $\lw{-1}(\cdot)$ the branch of the Lambert $W$ function \citep{Corless1996}
that satisfies $W(y) \leq -1$ for $y \in [-e^{-1}, 0)$, see Appendix
\ref{appendix:lambertW} for details about the Lambert $W$ function. The
sufficiently sceptical prior is then given by
$\theta \given \HS \sim \Nor(0, \gy \cdot \sigma^2_o)$ and it can be interpreted
as the view of a sceptic who argues that given their prior belief about the
effect $\theta$, the observed effect estimate $\hat{\theta}_o$ cannot convince
them about the presence of a non-null effect at level $\gamma$. An alternative
data-based interpretation of sufficiently sceptical priors is to see them as the
priors obtained by updating an initial uniform prior with the data from an
imaginary study, which was $1/\gy$ times the size of the original study, and
which resulted in an effect estimate of exactly zero \citep{Held2021b}.

From Figure \ref{fig:bfo} we can see that the more compelling the original data
(\ie the larger $|z_o|$), the smaller the sufficiently sceptical relative prior
variance $\gy$ needs to be in order to make the result no longer convincing at
level $\gamma$. In the most extreme case, when $|z_{o}| \to \infty$ and $\gamma$
remains fixed, the sufficiently sceptical prior variance will converge to
zero (Appendix \ref{appendix:lambertW}). On the other hand, if $|z_o|$ is not
sufficiently large, $\BFcol{0}{\text{S}}(\hat{\theta}_o; g)$ will either be
always increasing in $g$ (if $|z_o| \leq 1$) or it will reach a minimum above
the chosen level $\gamma$. In both cases the sufficiently sceptical relative
prior variance $\gy$ is not defined since there is no need to challenge an
already unconvincing result.

A side note on the Jeffreys-Lindley paradox is worth being mentioned: If a
$\gy < \gminBFo$ exists, there exists also a $\gy^\prime > \gminBFo$ as the
Bayes factor monotonically increases in $g > \gminBFo$ and therefore must
intersect a second time with $\gamma$, due to the paradox. This means that the
more compelling the original result, the larger $\gy^\prime$ needs to be chosen,
such that the result becomes no longer convincing at level $\gamma$. However,
priors which become increasingly diffuse do not represent increasing scepticism
but rather increasing ignorance. Using \eqref{ggamma} therefore avoids this
manifestation of the Jeffreys-Lindley paradox, since it determines sceptical
priors only from the class of priors that become increasingly concentrated for
increasing evidence (\ie priors with $\gy \leq \gminBFo$). In principle, the
solution $\gy^\prime > \gminBFo $ could also be computed by replacing the
$\lw{-1}$ branch of the Lambert $W$ function in \eqref{ggamma} with the $\lw{0}$
branch, but this will not be of interest to us.

\subsection{Data from the replication study}
\label{sec:bfr}
% In order to disprove the sceptic and confirm credibility of the original
% finding,
In order to assess whether the original finding can be replicated in an
independent study, a replication study is conducted, leading to a new effect
estimate $\hat{\theta}_r$. In light of the new data, the sceptic is now
challenged by an advocate of the original finding. This is formalised with
another Bayes factor, which compares the plausibility of the replication effect
estimate $\hat{\theta}_r \given \theta \sim \Nor(\theta, \sigma^2_r)$ under the
sceptical prior $\HS \colon$ $\theta \sim \Nor(0, g \cdot \sigma^2_o)$ relative
to the advocacy prior $\HA \colon$ $\theta \sim \Nor(\that_o, \sigma^2_o)$. The
the view of an advocate is represented by $\HA$ since this is the posterior of
$\theta$ given the original estimate and a uniform prior (also the reference
prior for this model). The Bayes factor is given by
\begin{align}
  \label{eq:BFr}
  \BFcol{\text{S}}{\text{A}}(\hat{\theta}_r; g)
%%   &= \sqrt{\frac{1/c + 1}{1/c + g}} \cdot
%%   \exp\left\{-\frac{1}{2}\left(\frac{z_r^2}{1/c + g} -
%%   \frac{\left(z_r - z_o \sqrt{c}\right)^2}{1 + c} \right)\right\}
  &= \sqrt{\frac{1/c + 1}{1/c + g}} \cdot
  \exp\left\{-\frac{z_o^2}{2}\left(\frac{d^2}{1/c + g} -
  \frac{\left(d - 1\right)^2}{1/c + 1} \right)\right\}
\end{align}
so it depends on the original $z$-statistic $z_o$, the relative sceptical prior
variance $g$, the relative effect estimate $d = \that_r/\that_o$, and the
relative variance $c = \sigma^2_o/\sigma^2_r$.

Our goal is now to define a condition for \emph{replication success} in terms of
\eqref{eq:BFr}. It is natural to consider a replication successful if the
replication data favour the advocate over the sceptic to a higher degree than
the sceptic's initial objection to the original study. More formally, we say
that if the Bayes factor from \eqref{eq:BFr} evaluated at the sufficiently
sceptical relative prior variance $\gy$ is not larger than the corresponding
level $\gamma$ used to define the sufficiently sceptical prior:
\begin{align}
  \label{eq:success}
  \BFcol{\text{S}}{\text{A}}(\hat{\theta}_r; \gy)
  \leq \BFcol{0}{\text{S}}(\hat{\theta}_o; \gy) =\gamma,
\end{align}
we have established \emph{replication success at level $\gamma$}.

For example, if we observe $z_o = 3$ (equivalent to minimum Bayes factor
$\minBF_{o} = 1/18$) and choose a level $\gamma = 1/10$
the sufficiently sceptical relative prior variance \eqref{ggamma} is
$\gy = 1.6$. If a replication is conducted with the
same precision ($c = 1$) and we observe $z_r = 2.5$ (equivalent to
minimum Bayes factor $\minBF_{r} = 1/5.5$ and relative effect
estimate $d = z_{r}/(z_{o}\sqrt{c}) = 0.83$), using equation
\eqref{eq:BFr} this would lead to $\BFcol{\text{S}}{\text{A}}(\hat{\theta}_r;
1.6) = 1/3.5$, which means
that the replication was not successful at level $\gamma = 1/10$.
However, if we had chosen a less stringent level, \eg$\gamma =
1/3$, the replication would have been considered successful
since then $\gy = 0.4$ and
$\BFcol{\text{S}}{\text{A}}(\hat{\theta}_r; 0.4)
= 1/7.4$.

\subsection{The sceptical Bayes factor}
Apart from specifying a level $\gamma$, the described procedure offers an
automated way to assess replication success. One way to remove this dependence
is to find the smallest level $\gamma$ where replication success can be
established. We thus call this level the \emph{the sceptical Bayes factor}
\begin{align}
  \label{eq:bfs}
  \BFs
  =
  %% \inf_{\gy \in [0, \gminBFo]}
  \inf
  \Big\{\gamma :
  \BFcol{\text{S}}{\text{A}}(\hat{\theta}_r; \gy)
  \leq \gamma\Big\},
\end{align}
and replication success at level $\gamma$ is equivalent with $\BFs \leq \gamma$.

\begin{figure}[!htb]
\begin{knitrout}
\definecolor{shadecolor}{rgb}{0.969, 0.969, 0.969}\color{fgcolor}
\includegraphics[width=\maxwidth]{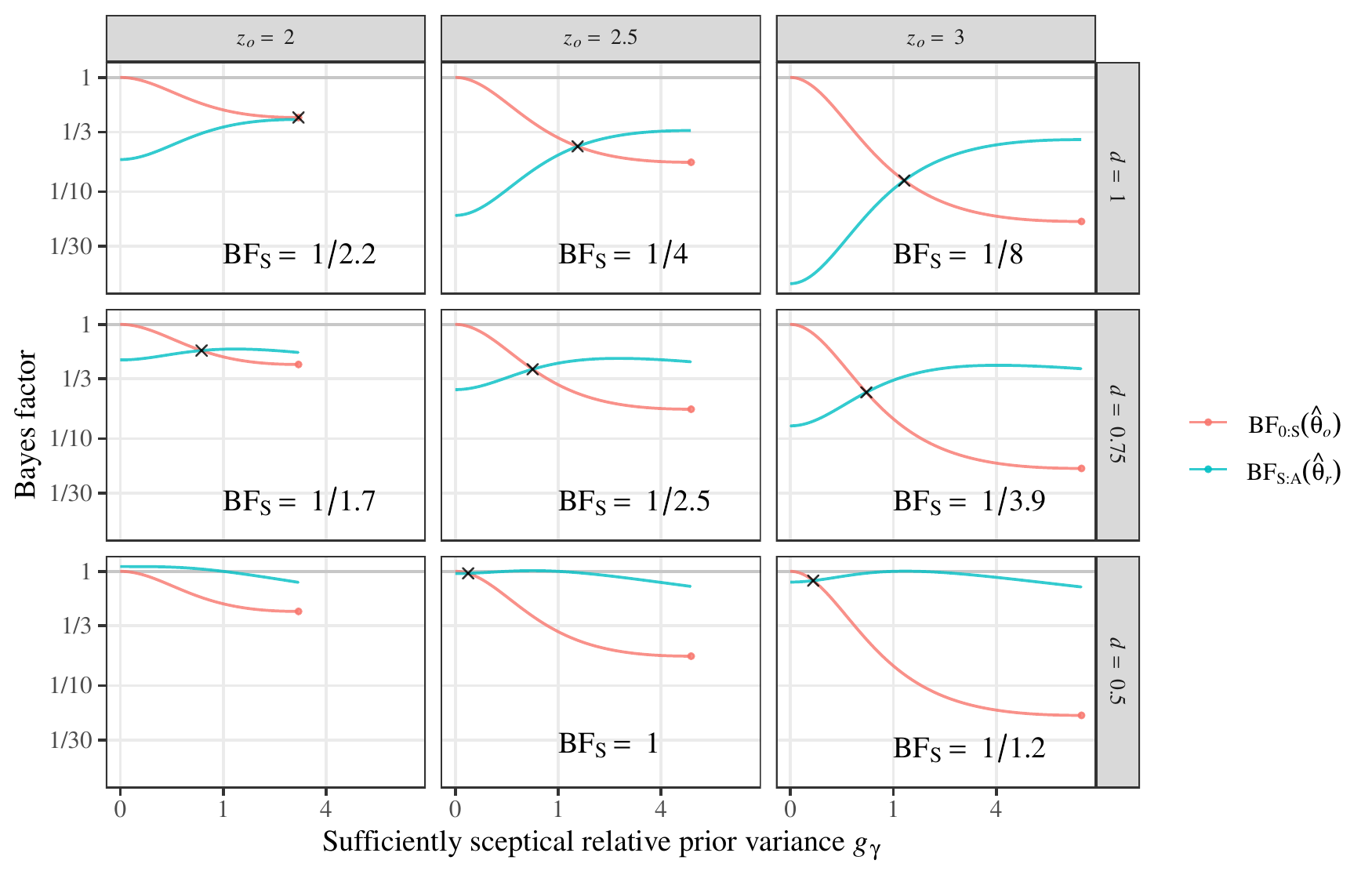}
\end{knitrout}
\caption{Bayes factors $\BFcol{\text{S}}{\text{A}}(\hat{\theta}_r; g)$ and
$\BFcol{0}{\text{S}}(\hat{\theta}_o; g)$ as a function of the sufficiently
sceptical relative prior variance $\gy$. In all examples
$c = \sigma^2_o/\sigma^2_r = 1$. Minimum Bayes factors $\minBF_o$ are indicated
by dots, sceptical Bayes factors $\BFs$ are indicated by crosses where existent.}
\label{fig:bfs}
\end{figure}

Figure \ref{fig:bfs} shows $\BFcol{\text{S}}{\text{A}}(\hat{\theta}_r; \gy)$ and
$\BFcol{0}{\text{S}}(\hat{\theta}_o; \gy)$ as a function of $\gy$ for several
values of $z_o$ and $d$ along with the corresponding $\BFs$. Typically, $\BFs$
is given by the height of the intersection between
$\BFcol{\text{S}}{\text{A}}(\hat{\theta}_r; \gy)$ and
$\BFcol{0}{\text{S}}(\hat{\theta}_o; \gy)$ in $\gy$. It may also happen that
$\BFcol{\text{S}}{\text{A}}(\hat{\theta}_r; \gy)$ remains below
$\BFcol{0}{\text{S}}(\hat{\theta}_o; \gy)$ for all values of $\gy$, in such
situations $\BFs$ is equal to the original minimum Bayes factor $\minBF_o$.
Finally, in some pathological cases it may happen that either $z_o$, $d$, or
both are so small that replication success cannot be established for any level
$\gamma$ and hence $\BFs$ does not exist. This means that the replication study
was unsuccessful since it is impossible for the advocate to convince the sceptic
at any level of evidence.

In terms of computing the sceptical Bayes factor, it is worth noting that for
the special case when the replication is conducted with the same precision as
the original study ($c = 1$) and $\BFs$ is located at the intersection of
$\BFcol{\text{S}}{\text{A}}(\hat{\theta}_r; g)$ and
$\BFcol{0}{\text{S}}(\hat{\theta}_o; g)$ in $g$, there is an explicit expression
for $\BFs$
\begin{align}
  \label{eq:BFsc1}
  \BFs =
  \sqrt{-\frac{z^2_{o}}{k} \cdot \frac{1 + d^{2}}{2}} \cdot
  \exp\left\{-\left(\frac{z_o^2}{2} + \frac{k}{1 + d^2}\right)\right\}
\end{align}
with
$$k = \lw{}\left(-\frac{z_{o}^{2}}{\sqrt{2}} \cdot \frac{d^{2} + 1}{2} \cdot \exp\left\{-\frac{z_o^2}{2}
  \left[1 + \frac{(1 -
d)^2}{2}\right]\right\}\right),$$ see Appendix \ref{appendix:bfs} for details.

\section{Properties}
\label{sec:comparison}
To study properties of the sceptical Bayes factor and facilitate comparison with
other methods we will look at the requirements for replication success based on
the relative effect estimate $d = \that_{r}/\that_{o}$, the variance ratio
$c = \sigma^{2}_{o}/\sigma^{2}_{r}$ and the original minimum Bayes factor
$\minBF_{o}$ (respectively the original $z$-value $z_{o}$). This perspective is
helpful because it disentangles how the method reacts to changes in
compatibility of the effect estimates ($d$), evidence from the original study
($\minBF_{o}$), and the change in sample size of the replication compared to the
original study ($c$).

The condition for replication success at level $\gamma$ from \eqref{eq:success}
is equivalent to
\begin{align}
  \label{eq:RSCond-d}
  \log \left\{ \frac{1/c + 1}{(1/c + \gy)(1 + \gy)}\right\} +
  \frac{z_o^2}{1 + 1/\gy} &\leq z_o^2 \left(\frac{d^2}{1/c +
    \gy} - \frac{\left(d - 1\right)^2}{1/c + 1} \right).
\end{align}
On the right-hand side of~\eqref{eq:RSCond-d} the $Q$-statistic
\begin{align}
  \label{eq:Qstat}
  Q
  = \frac{(\hat{\theta}_o - \hat{\theta}_r)^2}{\sigma^2_o + \sigma^2_r}
  %% = \frac{(z_r - z_o\sqrt{c})^2}{1 + c}
  = \frac{z_o^2(d - 1)^2}{1/c + 1}
\end{align}
appears. The $Q$-statistic was proposed as a measure of incompatibility among
original and replication effect estimates since its distribution is known for
standard meta-analytic models of effect sizes \citep{Hedges2019}. The connection
to the sceptical Bayes factor is such that $Q$ acts as a penalty term
in~\eqref{eq:RSCond-d} and a larger value will lower the degree of replication
success possible. However, as we will see, the sceptical Bayes factor goes
beyond assessing effect estimate compatibility as there is also a trade-off with
the amount of evidence that the replication study provides against the null.

Applying some algebraic manipulations to \eqref{eq:RSCond-d}, one can show that
replication success at level $\gamma$ is achieved if and only if the relative
effect estimate $d$ falls within a success region given by
\begin{align}
  \label{eq:dsuccess}
  \begin{cases}
    d \not\in [M - \sqrt{A/B}, M + \sqrt{A/B}] & \text{if} ~ \gy < 1 \\
    d \geq [1 + (1 + 1/c)\{1/2 - \log(2)/z_o^2\}]/2
    & \text{if} ~ \gy = 1 \\
   d \in [M - \sqrt{A/B}, M + \sqrt{A/B}] & \text{if} ~ \gy > 1\\
  \end{cases}
\end{align}
where
\begin{align*}
  M &= \frac{1/c + \gy}{\gy - 1} \\
  A &= \log \left\{ \frac{1/c + 1}{(1/c + \gy)(1 + \gy)}\right\}\Bigm/ z_o^2 +
  \frac{\gy}{1 + \gy} + \frac{1}{1 - \gy} \\
  B &= \frac{1 - \gy}{(1/c + \gy)(1/c + 1)}.
\end{align*}

\begin{figure}[!htb]
\begin{knitrout}
\definecolor{shadecolor}{rgb}{0.969, 0.969, 0.969}\color{fgcolor}
\includegraphics[width=\maxwidth]{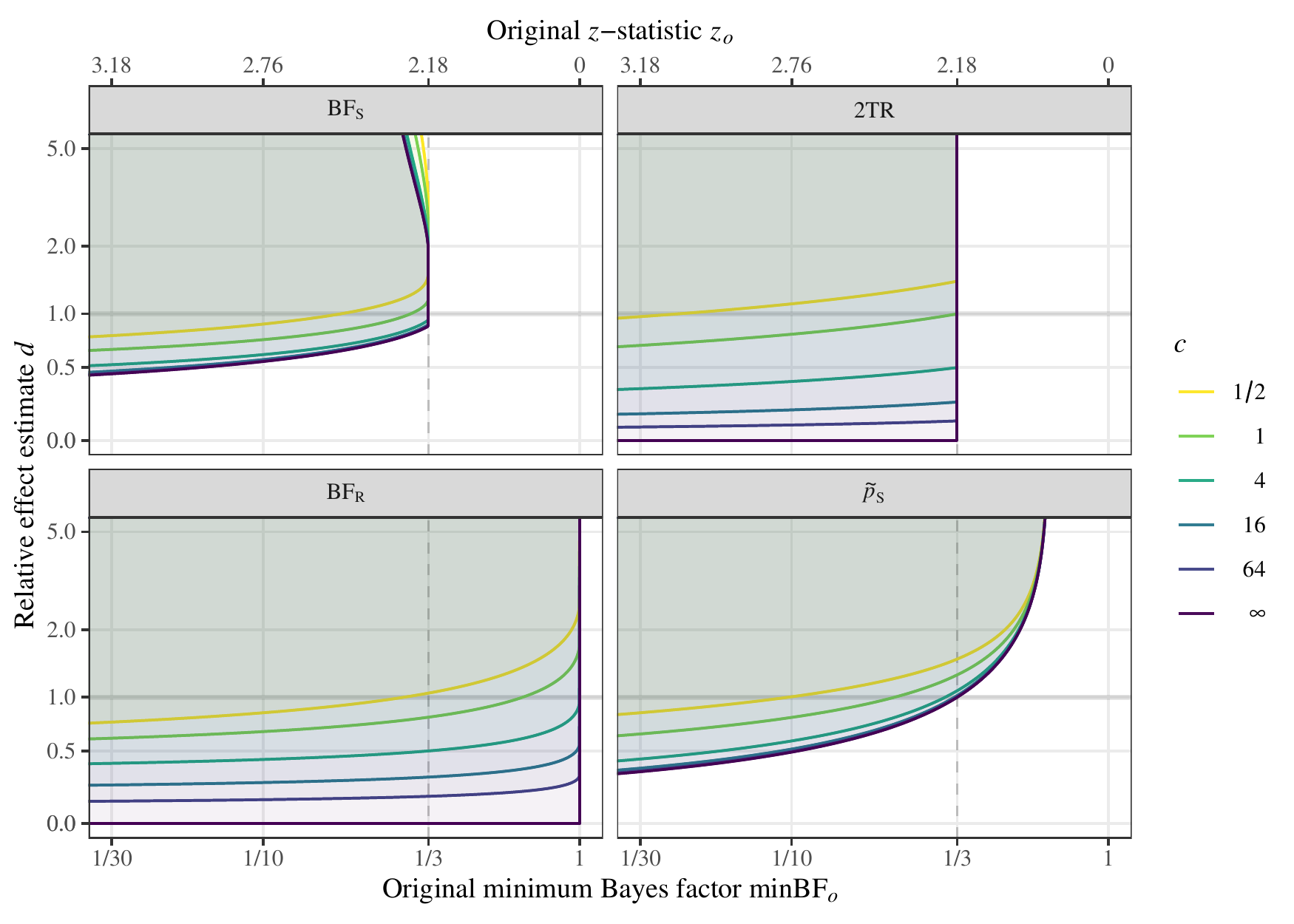}
\end{knitrout}
\caption{Required relative effect estimate $d = \that_{r}/\that_{o}$ to achieve
replication success based on the sceptical Bayes factor
($\BFs < 1/3$), the two-trials rule (2TR:
$\minBF_{o} < 1/3$ and $\minBF_{r} < 1/3$), the
replication Bayes factor ($\BFr < 1/3$), and the recalibrated
sceptical $p$-value ($\tilde{p}_{\text{S}} < 1 - \Phi\{z_{\gamma}\}$ with
$\gamma = 1/3$) as a function of the original minimum
Bayes factor $\minBF_{o}$ (respectively the corresponding $z$-value $z_{o}$) for
different values of the relative variance $c = \sigma^{2}_{o}/\sigma^{2}_{r}$.
Shading indicates regions where replication success is possible. Only positive
relative effect estimates $d$ are shown.}
\label{fig:dcomparison}
\end{figure}

The top-left plot in Figure~\ref{fig:dcomparison} shows the conditions on $d$
from~\eqref{eq:dsuccess} to achieve replication success at level $\gamma = 1/3$
as a function of the original minimum Bayes factor $\minBF_{o}$ and for
different values of the relative variance $c$. It is important to note that
$\gamma = 1/3$ is an arbitrary choice and in practice one should interpret the
sceptical Bayes factor as a quantitative measure of replication success. Only
the success regions for positive $d$ are shown as replication success in the
opposite direction is usually not of interest (see
Section~\ref{sec:replicationParadox} for a discussion of this issue). We see
that with increased precision of the replication study (larger $c$), the success
regions shift closer to zero. This means that the method allows for more
shrinkage of the replication effect estimate when the replication provides more
evidence against the null (because $\abs{z_{r}} = d \abs{z_{o}} \sqrt{c}$
increases with increasing $c$). However, the success regions cannot be pushed
arbitrarily close to zero but are bounded away. So when $c \to \infty$ the
methods still requires the replication estimate to be sufficiently large,
despite that the evidence against the null becomes overwhelming (since
$\abs{z_{r}} \to \infty$ as $c \to \infty$).

% level
By definition the sceptical Bayes factor can never be smaller than $\minBF_{o}$,
so replication success at level $\gamma$ is impossible for original studies with
$\minBF_{o} > \gamma$. This property is visible in the top-left plot in
Figure~\ref{fig:dcomparison} by the cut-off at $\minBF_{o} = \gamma = 1/3$. In
contrast, for more convincing original studies with $\minBF_{o} < 1/3$
replication success is possible and two cases can be distinguished in terms of
the success region: When $1/4.5 < \minBF_{o} \leq 1/3$ the
sufficiently sceptical relative prior variance is $\gy > 1$ and thus by
condition~\eqref{eq:dsuccess} the success region consists of an interval
$(\dmin, \dmax)$. Hence, in this case the method also penalises too large
replication effect estimates. For original studies with
$\minBF_{o} < 1/4.5$, the sufficiently sceptical relative prior
variance is $\gy \leq 1$, so due to~\eqref{eq:dsuccess} the success region for
positive $d$ is given by $(\dmin, \dmax = \infty)$. This means that for more
convincing original studies there are no upper restrictions for the relative
effect estimate, whereas shrinkage of the replication estimate is still
penalised.

\subsection{Comparison with other methods}
Of interest is the relationship between the sceptical Bayes factor and other
measures of replication success. Here, we review and compare a classical (the
two-trials rule), a forward-Bayes \citep[the replication Bayes factor
from][]{Verhagen2014} and a reverse-Bayes method \citep[the sceptical $p$-value
from][]{Held2020}. These methods provide a useful benchmark as they all are
based on hypothesis testing, have unique properties, and can be directly
compared in terms of their replication success regions as shown in
Figure~\ref{fig:dcomparison}.

\subsubsection{The two-trials rule}
Replication success is most commonly declared when both original and replication
study provide compelling evidence against a null effect. This approach is also
known as the \emph{two-trials rule} in drug development and usually a
requirement for drug approval \citep[Section 9.4]{Kay2015}. Most replication
projects report $p$-values associated with the effect estimates as measures of
evidence against the null, but also default Bayes factors have been used
\citep[see \eg the Bayesian supplement of][]{Camerer2018}. To compare the
two-trials rule with methods based on Bayes factors we will study the two-trials
rule based on the minimum Bayes factor from~\eqref{eq:minBFo}, \ie replication
success at level $\gamma$ is established when both
$\minBF_{k} % = \abs{z_{k}}\exp(-z_{k}^{2}/2)\sqrt{e}
< \gamma$ for $k \in \{o, r\}$, as well as
$\sign(\hat{\theta}_o) = \sign(\hat{\theta}_r)$. This approach has a one-to-one
correspondence to the usual version of two-trials rule as minimum Bayes factors
and $p$-values both only depend on the $z$-values of original and replication
study.

The two-trials rule guarantees that both studies provide compelling evidence
against the null. Similarly, the sceptical Bayes factor requires the original
study to be compelling on its own since it can never be smaller than
$\minBF_{o}$. However, one can easily construct examples where the sceptical
Bayes factor is smaller than the minimum Bayes factor from the replication study
(\eg when $\minBF_{o} = 1/2$,
$\minBF_{r} = 1/1.5$, and $c = 1$
we obtain $\BFs = 1/1.9$). So for the same level of
replication success $\gamma$ the two-trials may not flag replication success
whereas the sceptical Bayes factor would.

By definition the two-trials rule can never be fulfilled when the original study
was uncompelling. Assuming now that $\minBF_{o} < \gamma$, replication success
with the two-trials rule at level $\gamma$ is achieved if and only if the
relative effect estimate is
\begin{align}
  \label{eq:2TR}
  d \geq \frac{z_{\gamma}}{z_{o} \sqrt{c}}
\end{align}
with $z_{\gamma} > 1$ corresponding to
$\minBF = z_{\gamma}\exp(-z_{\gamma}^{2}/2)\sqrt{e} = \gamma$. The success
region from~\eqref{eq:2TR} is displayed in the top-right plot of
Figure~\ref{fig:dcomparison}. We see that the success regions shift closer to
zero as the relative variance $c$ increases. Also there is a cut-off at
$\minBF_{o} = \gamma = 1/3$ similarly as with the sceptical Bayes factor. In
contrast to the sceptical Bayes factor, however, the two-trials can be fulfilled
for any arbitrary small (but positive) relative effect estimate $d$, provided
the relative variance $c$ is large enough. Hence, the two-trials rule may flag
success even when the replication effect estimate is much smaller than the
original one.

\subsubsection{The replication Bayes factor}
\citet{Verhagen2014} proposed the \emph{replication Bayes factor} as a measure
of replication success. It is defined as the Bayes factor comparing the point
null hypothesis $H_0 \colon \theta = 0$, to the alternative that the effect is
distributed according to the posterior distribution of $\theta$ after observing
the original data. For the normal model considered so far and if an initial
reference prior was chosen, this alternative is also the advocacy prior
$\HA \colon \theta \sim \Nor(\that_o, \sigma^2_o)$ and therefore the replication
Bayes factor is given by
\begin{align}
  \label{eq:BFrep}
  \BFr
  = \BFcol{\text{0}}{\text{A}}(\hat{\theta}_r)
  %% = \sqrt{1 + c} \cdot
  %% \exp\left\{-\frac{1}{2} \left(z_r^2 -
  %% \frac{\left(z_r - z_o \sqrt{c}\right)^2}{1 + c} \right)\right\}.
  = \sqrt{1 + c} \cdot
  \exp\left\{-\frac{z_o^2}{2} \left(d^2 \cdot c -
  \frac{\left(1 - d\right)^2}{1/c + 1} \right)\right\}.
\end{align}

Similarly, as with the sceptical Bayes factor, the $Q$-statistic
from~\eqref{eq:Qstat} appears in~\eqref{eq:BFrep} and acts as a penalty term,
\ie larger values of $Q$ lower the degree of replication success. However, in
contrast to the sceptical Bayes factor, the replication Bayes factor is not
limited by the evidence from the original study because $\BFr \downarrow 0$ as
$c \to 0$ provided $z_{o} \neq 0$ and $d \neq 0$. Moreover, we have that
\begin{align*}
  1 \geq \BFs \geq \BFcol{\text{S}}{\text{A}}(\hat{\theta}_r; g_{\scriptscriptstyle \BFs})
  = \BFcol{\text{S}}{0}(\hat{\theta}_r; g_{\scriptscriptstyle \BFs}) \cdot \underbrace{\BFcol{0}{\text{A}}(\hat{\theta}_r)}_{= \BFr}.
\end{align*}
So the sceptical Bayes factor is larger than the replication Bayes factor if the
replication data favour the sceptical prior
$\HS \colon \theta \sim \Nor(0, g_{\scriptscriptstyle \BFs} \cdot \sigma^{2}_{o})$
over the null hypothesis. They can only coincide when $\BFs = 1$ since then
$g_{\scriptscriptstyle \BFs} = 0$.

We can also determine conditions on the relative effect estimate in terms of
replication success based on $\BFr < \gamma$. The replication success region is
given by
\begin{align}
  \label{eq:BFRsuccess}
  d \not\in (-\sqrt{K} - H, \sqrt{K} - H)
\end{align}
with
\begin{align*}
  K &= \left\{1 + \frac{\log(1 + c) - 2 \log \gamma}{z_o^2}\right\} \cdot
  \frac{1/c + 1}{c} \\
  H &= \frac{1/c + 1}{1 + c}.
\end{align*}
The condition \eqref{eq:BFRsuccess} implies that replication success can also be
achieved for negative relative effect estimates $d$ (see
Section~\ref{sec:replicationParadox} for a discussion of this issue). The
bottom-left plot in Figure~\ref{fig:dcomparison} shows the conditions
from~\eqref{eq:BFRsuccess} for positive relative effect estimates. As with the
two-trials rule, the success region of the replication Bayes factor can be
pushed arbitrarily close to zero by increasing the relative variance $c$. In
contrast to the two-trials rule, however, replication success can also be
achieved for original studies with $\minBF_{o} > 1/3$.

\subsubsection{The sceptical $p$-value}
Of particular interest is the relationship between the sceptical Bayes factor
and the sceptical $p$-value \citep{Held2020}, as it is the outcome of a similar
reverse-Bayes procedure. One also considers a sceptical prior for the effect
size $\theta \sim \Nor(0, \tau^2)$, the sufficiently sceptical prior variance at
level $\alpha$ is then defined as $\tau^2 = \tau^2_\alpha$ such that the
$(1 - \alpha)$ credible interval for $\theta$ based on the posterior
$\theta \given \that_o, \tau^2_\alpha$ does not include zero. Replication
success is declared if the tail probability of the replication effect estimate
under its prior predictive distribution
$\that_r \given \tau^2_\alpha \sim \Nor(0, \tau^2_\alpha + \sigma^2_r)$ is
smaller than $\alpha$. The smallest level $\alpha$ where replication success can
be established defines the sceptical $p$-value. In contrast to the sceptical
Bayes factor, the sceptical $p$-value always exists and there are closed form
expressions to compute it for all values of $c$, \ie
$\ps = 1 - \Phi(z_{\text{S}})$ with
\begin{equation*}
  z^2_{\text{S}} =
  \begin{cases}
    z^2_H/2 & \text{for}~c=1 \\ \frac{1}{c - 1}\left\{\sqrt{z^2_A\left[z^2_A +
          (c - 1)z^2_H\right]} - z^2_A\right\} & \text{for}~c \neq 1
  \end{cases}
  \label{eq:z2s}
\end{equation*}
where $z^2_H = 2/(1/z_o^2 + 1/z_r^2)$ the harmonic mean,
$z^2_A = (z_o^2 + z_r^2)/2$ the arithmetic mean of the squared $z$-statistics,
and provided that $\sign(\that_o) = \sign(\that_r)$ (otherwise
$\ps = \Phi(z_{\text{S}})$).

Similar to the two-trials rule, the sceptical $p$-value requires both studies to
provide compelling evidence due to the property that
$\ps \geq \max\{p_o, p_r\}$. The sceptical $p$-value also penalises the case
when the replication effect estimate shrinks as compared to the original one
since it monotonically increases with decreasing relative effect estimate $d$
\citep[Section 3.1]{Held2020}.

\citet{Held2021} showed that that replication success based on
$\ps \leq \alpha_{\text{S}}$ is achieved when
\begin{align}
  \label{eq:psSuccessd}
  d \geq \sqrt{\frac{1/c + 1/(K - 1)}{K}}
\end{align}
with $K = z_{o}^{2}/z_{\alpha_{\text{S}}}^{2}$ where
$z_{\alpha_{\text{S}}} = \Phi^{-1}(1 - \alpha_{\text{S}})$. Thresholding the
sceptical $p$-value with the ordinary significance level $\alpha$ for
traditional $p$-values leads to a very stringent criterion for replication
success. For example, when $z_{o} = 2$, $\alpha = 0.025$, and
$c = 2$, the replication effect estimate needs to be
$d = 4.87$ times larger than the original one. Therefore,
\citet{Held2021} used~\eqref{eq:psSuccessd} to determine the \emph{golden level}
$\alpha_{\text{S}} = 1 - \Phi(z_\alpha/\sqrt{\varphi})$ with
$\varphi = (1 + \sqrt{5})/2$ the golden ratio. The golden level ensures that
borderline significant original studies ($\abs{z_{o}} = z_{\alpha}$) can still
achieve replication success provided the replication effect estimate does not
shrink compared to the original one ($d \geq 1$). Instead of comparing the
sceptical $p$-value to the golden level ($\ps < \alpha_{\text{S}}$), one can
compute a recalibrated sceptical $p$-value
$\tilde{p}_{\text{S}} = 1 - \Phi(z_{\text{S}} \sqrt{\varphi})$ and compare it to
the ordinary significance level ($\tilde{p}_{\text{S}} < \alpha$).

The bottom-right plot in Figure~\ref{fig:dcomparison} shows the success region
for the recalibrated sceptical $p$-value. We see that increasing the precision
of the replication study lowers the required minimum relative effect estimate
$\dmin$ as for all other methods. Similarly, as with the sceptical Bayes factor,
$\dmin$ of the sceptical $p$-value cannot be pushed arbitrarily close to zero.
However, its limiting minimum relative effect estimate in $c$
($\lim_{c \to \infty}\dmin$) is smaller than the one from the sceptical Bayes
factor when $\minBF_{o} < 1/5.6$, while for
$\minBF_{o} > 1/5.6$ it is the other way around. So for more
convincing original studies the sceptical $p$-value is less stringent than the
sceptical Bayes factor. Due to the recalibration, the sceptical $p$-value also
allows replication success when the $\minBF_{o} > \gamma$. This is visible in
the bottom-right plot of Figure~\ref{fig:dcomparison} where the success region
has no cut-off at $\minBF_{o} = \gamma = 1/3$, unlike the two-trials rule and
the sceptical Bayes factor.

\subsection{Paradoxes in the assessment of replication success}
The replication setting is different from the classical setting where data from
only one study are analysed. As a result, several unique paradoxes may occur.

\subsubsection{The replication paradox}
\label{sec:replicationParadox}
The \emph{replication-paradox} \citep{Ly2018} occurs when original and
replication effect estimates go in opposite directions
($\sign(\that_{o}) \neq \sign(\that_{r})$) but a method flags replication
success. This is undesired since effect direction is crucial to most
scientific theories and research questions.

The two-trials rule and the sceptical $p$-value both avoid the replication
paradox by using one-sided test-statistics. In contrast, the sceptical Bayes
factor and the replication Bayes factor may suffer from the paradox as their
success regions from~\eqref{eq:dsuccess} and~\eqref{eq:BFRsuccess},
respectively, include negative relative effect estimates $d < 0$. This is
related to the fact that Bayes factors are quantifying relative evidence: When
the replication estimate goes in the opposite direction, it will be poorly
predicted by the sceptical prior $\HS$ and the advocacy prior $\HA$, yet when
$\HS$ is mostly concentrated around zero (or a point-null in case of the
replication Bayes factor), replication estimates going in the opposite direction
may still be better predicted by $\HA$.

In practice, the replication paradox is hardly an issue, since replications
rarely show such contradictory results, \eg to achieve replication success at
level $\gamma = 1/3$ with
$\minBF_{o} = 1/10$ and $c = 1$, the
relative effect estimate needs to be $d < -7.09$ for the
paradox to appear with the sceptical Bayes factor. The replication Bayes factor
is more prone to the paradox because its point-null hypothesis fails more
strongly to predict estimates in opposite direction, \eg for the same numbers as
before it requires $d < -2.66$.

In both cases the paradox can be overcome by truncating the advocacy prior $\HA$
such that only effects in the same direction as the original estimate $\that_o$
have non-zero probability, \ie for positive $\that_{o}$ consider
$\HAp \colon \theta \sim \Nor(\that_o, \sigma^2_o) \, \mathbb{1}_{(0, \infty)}(\theta)$,
where $\mathbb{1}_{B}(x)$ is the indicator function of the set $B$. The Bayes
factor contrasting $\HS$ to $\HAp$ turns out to be
\begin{align}
  \BFcol{\text{S}}{\text{A}^\prime}(\hat{\theta}_r; g)
  &= \BFcol{\text{S}}{\text{A}}(\hat{\theta}_r; g)
  \frac{\Phi(\abs{z_o})}{\Phi\left\{\text{sign}(z_o)
    %     \frac{z_o + z_r\sqrt{c}}{\sqrt{1 + c}} \right\}}
    \frac{z_o(1 + dc)}{\sqrt{1 + c}} \right\}}
   \label{eq:BFrtrunc}
\end{align}
where $\Phi(\cdot)$ is the cumulative distribution function of the standard
normal distribution (see Appendix \ref{app:BFtrunc}). Hence, \eqref{eq:BFrtrunc}
is the Bayes factor under the standard advocacy prior multiplied by a correction
term. Determining the smallest level of replication success
with~\eqref{eq:BFrtrunc} leads to a corrected sceptical Bayes factor, while
setting $g = 0$ in~\eqref{eq:BFrtrunc} leads to a corrected replication Bayes
factor. The correction term goes to one when the replication estimate goes in
the same direction as the original one and the replication sample size increases
($d > 0$ and $c \to \infty$), but it penalises when the replication estimate
goes in the opposite direction ($d < 0$ and $c \to \infty$).

This modification guarantees that the replication paradox is avoided and we
recommend to compute the sceptical Bayes factor using~\eqref{eq:BFrtrunc} in
cases where the replication paradox is likely to appear. However, truncated
priors are unnatural and hard to interpret. Also the non-truncated advocacy
prior penalises effect estimate incompatibility and the
modification~\eqref{eq:BFrtrunc} will only make a difference in extreme
situations. Due to its easier mathematical treatment we will focus on the
standard version of the procedure in the remaining part of the manuscript.

\subsubsection{The shrinkage paradox}
The comparison showed that for certain methods replication success is still
achievable even when the replication estimate is substantially smaller than the
original one. However, a substantially smaller effect estimate in the
replication does not reflect an effect size of the same practical importance as
the original one and a method should thus not flag replication success. The {\em
shrinkage paradox} occurs if a particular method may flag replication success
for any arbitrarily small (but positive) relative effect estimate.

Two forms of the shrinkage paradox can formally be distinguished: the
\emph{shrinkage paradox at replication} appears when, for fixed evidence from
the original study $\minBF_{o}$ (respectively $z_{o}$), the minimum relative
effect estimate $\dmin > 0$ required for replication success at a fixed level
$\gamma$ becomes arbitrarily small as the relative variance $c$ increases:
\begin{align*}
  \dmin \downarrow 0 ~\text{as}~ c \to \infty.
\end{align*}
\citet{Held2021} found that this form the paradox occurs for the two-trials rule
but not for the sceptical $p$-value. Similarly, the minimum relative effect
estimate $\dmin$ of the sceptical Bayes factor is bounded away from zero, while
it converges to zero for the replication Bayes factor
(Appendix~\ref{app:shrinkage}). Hence, among the Bayesian methods, the sceptical
Bayes factor avoids the paradox, whereas the replication Bayes factor suffers
from it.

The shrinkage paradox at replication is a serious issue since it depends on the
relative variance $c$ which can usually be directly influenced by changing the
replication sample size. However, there is also a second form of the paradox
which is affected only by evidence from the original study. The \emph{shrinkage
paradox at original} appears when, for fixed relative variance $c$, the minimum
relative effect estimate $\dmin > 0$ required for replication success at a fixed
level $\gamma$ becomes arbitrarily small as the evidence in the original study
increases:
\begin{align*}
  \dmin \downarrow 0 ~\text{as}~ z_{o}^{2} \to \infty.
\end{align*}

The replication Bayes factor and the sceptical Bayes factor do not suffer from
this form of the paradox, while the two-trials rule and the sceptical $p$-value
do (Appendix~\ref{app:shrinkage}). Hence, with the latter two methods, shrinkage
of the replication effect estimate is hardly penalised when the original study
was already very convincing.

\subsection{Frequentist properties}
Despite the fact that Bayesian methods do not rely on repeated testing, it is
still often of interest to study their frequentist operating characteristics
\citep{Dawid1982, Grieve2016}. This is especially important in the replication
setting where regulators and funders usually require from statistical methods to
have appropriate error control. We will therefore study and compare type I error
rate as well as power of the sceptical Bayes factor and other methods.

\subsubsection{Global type I error rate}
The probability for replication success at level $\gamma$ conditional on the
original result $z_{o}$ and the relative variance $c$ can be easily computed as
shown in Appendix~\ref{app:powerBFs}. Under the null hypothesis
($H_{0}: \theta = 0$) the distribution of the $z$-values is
$z_{o}, z_{r} \given H_{0} \sim \Nor(0, 1)$ and hence the global type I error
rate (T1E) based on $\BFs \leq \gamma$ is
\begin{align*}
  \text{T1E} = 2 \int_{z_{\gamma}}^{\infty} \P(\BFs \leq \gamma \given z_{o}, c) \phi(z_{o}) \,
  \text{d}z_{o}
\end{align*}
with $\phi(\cdot)$ the standard normal density function. In a similar fashion
one can compute the type I error rate of the sceptical $p$-value \citep[see
Section 3 in][]{Held2021}, as well as the replication Bayes factor
(Appendix~\ref{app:powerBFr}). The type I error rate of the two trials rule is
simply $\text{T1E} = 2\{1 - \Phi(z_{\gamma})\}^{2}$.

\begin{figure}[!htb]
\begin{knitrout}
\definecolor{shadecolor}{rgb}{0.969, 0.969, 0.969}\color{fgcolor}
\includegraphics[width=\maxwidth]{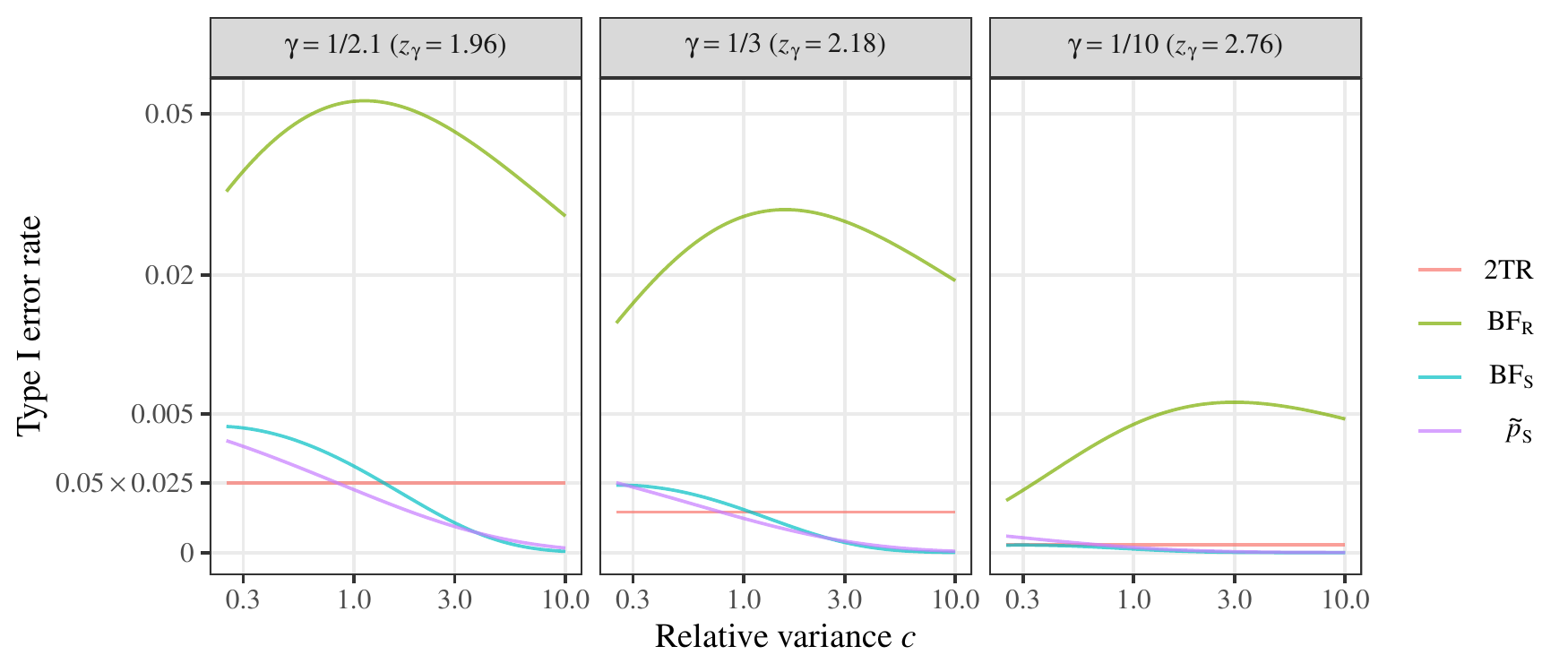}
\end{knitrout}
\caption{Type I error rate of the two-trials rule (2TR: $\minBF_{o} < \gamma$
and $\minBF_{r} < \gamma$), the replication Bayes factor ($\BFr < \gamma$), the
sceptical Bayes factor ($\BFs < \gamma$), and the recalibrated sceptical
$p$-value ($\tilde{p}_{\text{S}} < 1 - \Phi\{z_{\gamma}\}$) as a function of the
relative variance $c = \sigma^{2}_{o}/\sigma^{2}_{r}$ for different levels
of replication success $\gamma$.}
\label{fig:t1ec}
\end{figure}

Figure~\ref{fig:t1ec} compares the type I error rates of the four methods for
different levels $\gamma$. The conventional nominal
$\text{T1E} = 0.05 \times 0.025$ (two independent experiments with two-sided
testing in the first and one-sided testing in the second) along with the
corresponding level ($z_{\gamma} = 1.96$ corresponding to $\gamma = 1/2.1$ and
$\alpha = 0.025$) is also indicated. In contrast to the other methods, the type
I error rate of the two trials rule does not depend on the relative variance $c$
and therefore does not change for the same level $\gamma$. Type I error rates of
sceptical $p$-value and sceptical Bayes factor are decreasing with increasing
$c$, the former usually being slightly smaller than the latter. The point at
which both become smaller than the type I error rate from the two-trials rule
becomes smaller with more stringent level $\gamma$. Roughly speaking the type I
error rate of the sceptical Bayes factor is controlled at the conventional level
when $c$ is slightly larger than one, while for the sceptical $p$-value it is
controlled when $c$ is slightly below one. Surprisingly, the type I error rate
of the replication Bayes factor is non-monotone in $c$ and far higher compared
to the other methods. This suggests that a more stringent level $\gamma$ should
be used for the replication Bayes factor compared to the other methods to ensure
appropriate type I error control.

\subsubsection{Power conditional on the original study}
Another frequentist operating characteristic is the probability to establish
replication success assuming there is an underlying effect (power). While in
principle original and replication study could be powered simultaneously, we
will assume the original study has already been conducted since this is the
usual situation. The power to establish replication success $\BFs \leq \gamma$
can be computed using the result from Appendix~\ref{app:powerBFs} and either
assuming that the underlying true effect corresponds to its estimate from the
original study (\emph{conditional power}) or using the predictive distribution
of the replication effect estimate based on the advocacy prior (\emph{predictive
power}) \citep{Spiegelhalter1986b, Micheloud2021}. In practice, both forms may
be too optimistic as original results are often inflated due to publication bias
and questionable research practices. One solution is to shrink the original
effect estimate for power calculations \citep{Pawel2020, Held2021}, but we will
not focus on this aspect here as this would not provide much more insight but
simply lower the power curves of all methods.

\begin{figure}
\begin{knitrout}
\definecolor{shadecolor}{rgb}{0.969, 0.969, 0.969}\color{fgcolor}
\includegraphics[width=\maxwidth]{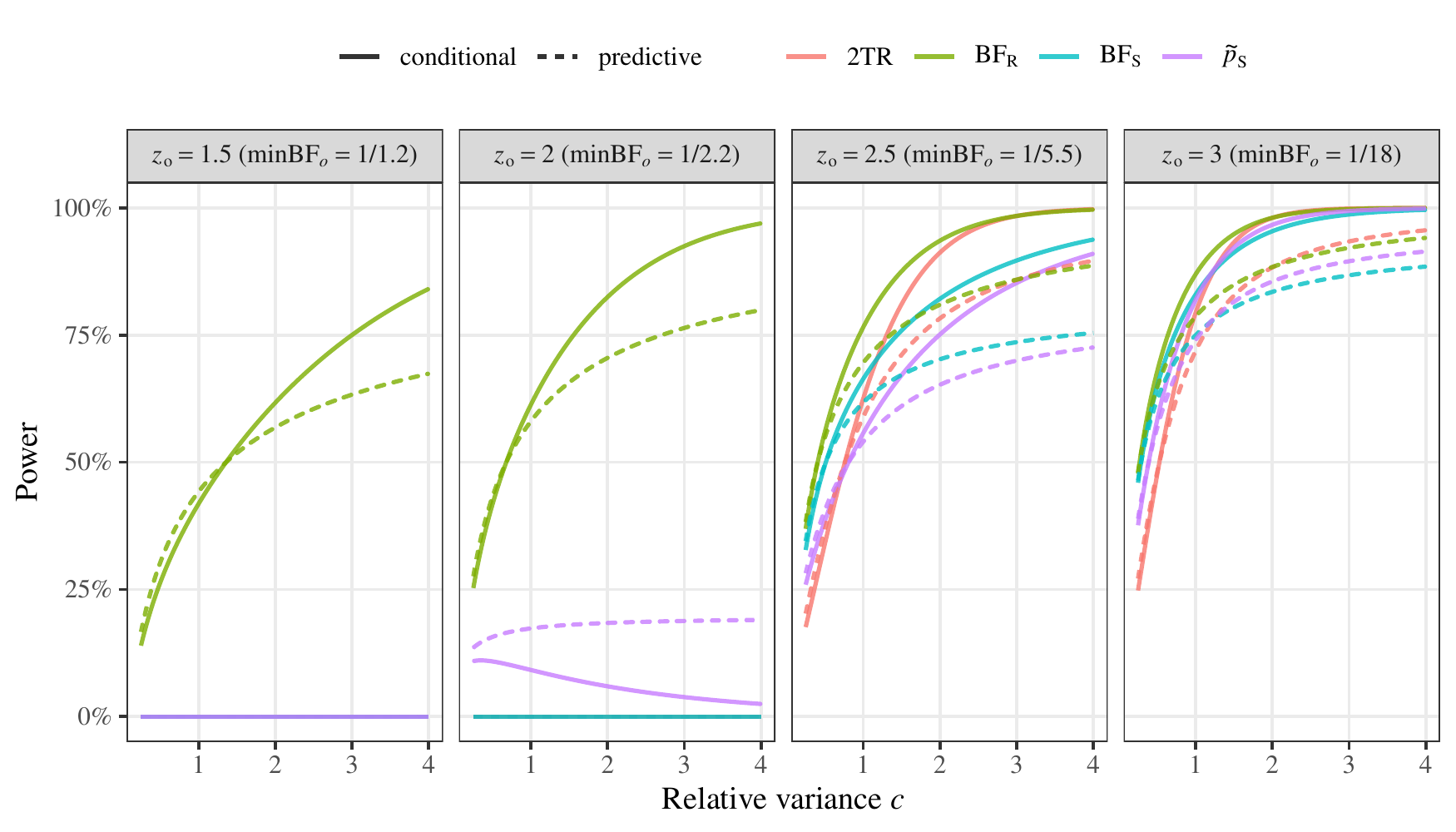}
\end{knitrout}
\caption{Power of the two-trials rule (2TR: $\minBF_{o} <
1/3$ and $\minBF_{r} <
1/3$), the replication Bayes factor ($\BFr <
1/3$), the sceptical Bayes factor ($\BFs <
1/3$), and the recalibrated sceptical $p$-value ($\tilde{p}_{\text{S}}
< 1 - \Phi\{z_{\gamma}\}$ with $\gamma =
1/3$) as a function of the relative variance $c
= \sigma^2_o/\sigma^2_r$ for different original $z$-values $z_o = \that_o/\sigma_o$
(respectively corresponding minimum Bayes factor $\minBF_o$).}
\label{fig:condpower}
\end{figure}

Figure~\ref{fig:condpower} shows conditional and predictive power as a function
of the relative variance $c$ and for several values of of the original $z$-value
$z_o$ (respectively original minimum Bayes factor $\minBF_{o}$). In general,
uncertainty about replication success is higher for predictive power, leading it
to be closer to 50\% in all cases. As can also be seen, if the original result
was not convincing on its own (\eg if $z_o = 1.5$ or $z_{o} = 2$), it is
impossible to achieve replication success with the two-trials rule, the
sceptical Bayes factor, and the sceptical $p$-value. This is not the case for
the replication Bayes factor, for which high power can also be obtained for
small $z_o$ if $c$ is sufficiently large. However, as shown in the previous
section, the higher power of the replication Bayes factor comes at the cost of a
massive type I error inflation. For $z_o = 2.5$, the sceptical Bayes factor
shows higher power than the the two-trials rule when $c = 1$, but the power of
the two-trials rule increases faster in $c$ and approaches the power curve of
the replication Bayes factor. The power of the sceptical $p$-value is still a
bit lower, likely due to the more stringent requirement on the minimum relative
effect estimate. For $z_{o} = 3$, the power differences between the methods
mostly disappear.
% This is
% expected, since the power of the sceptical Bayes factor and the replication
% Bayes factor should be similar when the sufficiently sceptical prior is strongly
% concentrated concentrated around zero.

\subsection{Information consistency}
\label{sec:informationcons}
Bayesian hypothesis testing procedures are desired to fulfil certain asymptotic
properties \citep{Bayarri2012}. Most notably, they should be \emph{information
consistent} in the sense that if data provide overwhelming support for a
particular hypothesis, the procedure should indefinitely favour this hypotheses
over alternative hypotheses.
% Whether or not a procedure is information consistent is usually investigated by
% either letting the sample size or a summary statistic of the data go to infinity.
% For example, \citet{Liang2008} studied consistency of
% Bayes factors in a regression setting by letting the Euclidean norm of
% the regression coefficients go to infinity for a fixed sample size, or
% \citet{Jeffreys1961} studied consistency of Bayes factors for testing
% the mean of a normal distribution by letting the sample size go to infinity
% and fixing the data mean.

There are concerns whether the sceptical Bayes factor is information consistent
when we look at the asymptotics only in terms of the replication data
\citep{Consonni2021, Ly2021}. The sceptical Bayes factor can never be smaller
than the original minimum Bayes factor $\minBF_{o}$. This means that it will be
bounded away from zero as the replication sample size grows ($c \to \infty$),
even when the data are generated from the same model in both studies. Similarly,
the sceptical Bayes factor will be bounded away from zero when the replication
effect estimate increases indefinitely. If these two cases constitute
overwhelming evidence for replication success, they could be considered
instances of the \emph{information paradox} \citep{Liang2008}
% Similarly, with increasing replication effect estimate
% ($\that_{r} \to \infty$ implying that also the relative effect estimate
% $d \to \infty$ when $\that > 0$), replication success can at most be established for a
% sufficiently sceptical relative prior variance $\gy \leq 1$ due to the
% success conditions~\eqref{eq:dsuccess}. This implies that the sceptical Bayes factor
% $\BFs \to \BFcol{0}{\text{S}}(\hat{\theta}_o; g = 1)$, provided $\abs{z_o} \geq
% \sqrt{2}$ so that $g_{\minBF_{o}} \geq 1$. Hence, the sceptical Bayes factor is
% also bounded away from zero and this could be considered an
% instance of the \emph{information paradox} \citep{Liang2008}.

The key to resolving the paradox is to realise that overwhelming evidence for
replication success needs to be defined through both studies, not only through
the replication. Assume there is a ``true'' effect size $\theta_{*} \neq 0$
underlying both effect estimates
$\that_{i} \sim \Nor(\theta_{*}, \sigma_{i}^{2})$, $i \in \{o, r\}$. Also assume
the variances $\sigma_{i}^{2} = \kappa/n$ are inversely proportional to the
sample size $n = n_{o} = n_{r}$ for the same unit variance $\kappa$ in both
studies. Letting the sample size $n$ go to infinity is then equivalent to
$\sigma_{o}^{2} \downarrow 0$ and $c = \sigma^{2}_{r}/\sigma^{2}_{o} = 1$. With
decreasing variances the estimates will
converge to the true effect size ($\that_{i} \to \theta_{*}$), the relative
effect estimate will converge to one ($d \to 1$), and the $z$-values will go to
infinity ($\abs{z_{i}} \to \infty$). Since $c = 1$, the sceptical Bayes factor
is given by equation~\eqref{eq:BFsc1}. Moreover, we are allowed to use of the
approximation $\lw{-1}(x) \approx \log(-x) - \log(-\log(-x))$ as the argument of
the Lambert function is close to zero due to $\abs{z_{o}} \to \infty$ \citep[p.
350]{Corless1996}. Taken together, we have
\begin{align}
  \label{eq:BFsc1lim2}
  \BFs =
  \sqrt{\frac{1 + d^{2}}{1 + (1 - d)^{2}/2 -
\mathcal{O}\left\{\log(z_{o}^{2})/z_{o}^{2}\right\}}} \cdot
  \exp\left\{-\frac{z_o^2\left(d^{2} + 2d - 1 \right)}{4(1 + d^{2})}
- \mathcal{O}\left(\log z_{o}^{2}\right)\right\}
\end{align}
Plugging $d = 1$ into~\eqref{eq:BFsc1lim2}, we see that $\BFs \downarrow 0$ as
$\abs{z_{o}} \to \infty$, so the sceptical Bayes factor is information
consistent.

The expression for the sceptical Bayes factor~\eqref{eq:BFsc1lim2} is also valid
for other relative effect sizes $d$. Solving for $d$ such that the
multiplicative term of $z_{o}^{2}$ in the exponent changes the sign, we see that
the sceptical Bayes factor goes to zero when the underlying true effect size of
the replication study is at least
$d > \sqrt{2} - 1 \approx 0.41$ times the size of the
true effect size from the original study (or
$d < -\sqrt{2} - 1 \approx -2.41$ due to the
replication paradox if the advocacy prior is not truncated). This means that
under the more realistic scenario where the underlying effect sizes from
original and replication are not exactly the same, the sceptical Bayes factor is
still consistent when there is not more than 60\% shrinkage of the replication
effect size.

\section{Extension to non-normal models} \label{sec:tdist}
So far, we have always assumed approximate normality of the effect estimates
$\that_o$ and $\that_r$, as well as known variances $\sigma^2_o$ and
$\sigma^2_r$. This may be a problem for studies with small sample size and/or
extreme results (\eg when a study examines a rare disease with death rates close
to 0\%). One way of dealing with this issue is to consider the exact likelihood
of the data underlying the effect estimates, and then marginalise over possible
nuisance parameters \citep[Chapter 8.2.2]{Spiegelhalter2004}. This leads to
marginal likelihoods which are again only conditional on the effect size
$\theta$, allowing the procedures to be used analogously as described in the
proceeding sections. The choice of the likelihood depends on the type of effect
size $\theta$. We will illustrate the approach for \emph{standardised mean
differences} (SMD) and \emph{log odds ratios} (logOR), two of the most widely
used types of effect sizes.

\subsection{Standardised mean difference}
The SMD quantifies how many standard deviation units $\sigma$, the means $\mu_1$
and $\mu_2$ of measurements from two groups differ, \ie
$$\theta = \frac{\mu_1 - \mu_2}{\sigma}.$$
Assume now that the measurements come from a normal distribution with common
variance $\sigma^2$. Knowing the test-statistic $t_i$ from the usual two-sample
$t$-test, as well as the sample sizes in both groups $n_{1i}$ and $n_{2i}$ from
study $i \in \{o, r\}$ is sufficient to compute the exact likelihood of the
data. It is given by a non-central $t$-distribution with degrees of freedom
$\nu_i = n_{1i} + n_{2i} - 2$ and non-centrality parameter $\theta \sqrt{n^*_i}$
with $n^*_i = (n_{1i}n_{2i})/(n_{1i} + n_{2i})$ \citep{Bayarri2002}
\begin{align}
  \label{eq:nct}
  T_i \given \theta \sim \text{NCT}_{\nu_i} \left(\theta \sqrt{n^*_i}\right).
\end{align}
The same framework is also applicable to test-statistics $t_{i}$ from paired
$t$-tests based on $n_{i}$ paired measurements. The SMD $\theta$ represents then
the standardised mean difference score and $\nu_{i} = n_{i} - 1$ and
$n_{i}^{*} = n_{i}$ need to be used in~\eqref{eq:nct}.

There is no conjugate prior for the SMD $\theta$ under model~\eqref{eq:nct}, so
it is not obvious which prior should be chosen to represent scepticism about it.
We will use a zero-mean normal prior $\theta \given \HS \sim \Nor(0, \tau^{2})$
so that the exact procedure is equivalent with the normal approximation as the
sample size increases. For the advocacy prior we need to know the posterior
distribution of the SMD $\theta$ conditional on the original study and a flat
prior on $\theta$. Exploiting the fact that the non-central $t$-distribution can
be expressed as a location-scale mixture of a normal with an inverse-gamma
distribution
%% \ie $\text{NCT}_\nu(t; \theta \sqrt{n^*}) = \int
%% \Nor(t; \theta \sqrt{n^*\tau^2}, \tau^2) \text{IG}(\tau^2; \nu/2, \nu/2) d\tau^2$
\citep[Chapter 31]{Johnson1995}, the density of the SMD under the advocacy prior
is given by
\begin{align*}
  f(\theta \given t_o)
  = \int_0^\infty \Nor\left(\theta; \frac{t_o}{\sqrt{n_o^* \tau^2}},
  \frac{1}{n_o^*}\right)  \text{IG}\left(\tau^2; \frac{\nu_o + 1}{2},
  \frac{\nu_o}{2}\right) \, \text{d}\tau^2,
\end{align*}
where $\Nor(x; \mu, \phi)$ denotes the density function of the normal
distribution with mean $\mu$ and variance $\phi$ evaluated at $x$, and
similarly $\text{IG}(y; a, b)$ denotes the density function of the inverse-gamma
distribution with parameters $a$ and $b$ evaluated at $y$.

Taken together, the SMD version of the method proceeds analogously as in
Box~\hyperref[box:nutshell]{1} with the two Bayes factors replaced by
\begin{align*}
  \BFcol{0}{\text{S}}(t_{o};\tau^{2})
  &= \frac{\text{NCT}_{\nu_{o}}(t_{o}; 0)}{\int
  \text{NCT}_{\nu_{o}}(t_{o}; \theta \sqrt{n_{o}^{*}})
  \Nor(\theta; 0, \tau^{2}) \, \text{d}\theta} \\
  \BFcol{\text{S}}{\text{A}}(t_{r};\tau^{2})
  &= \frac{\int\text{NCT}_{\nu_{r}}(t_{r};\theta
  \sqrt{n_{r}^{*}})\Nor(\theta; 0, \tau^{2}) \, \text{d}\theta}{
  \int\text{NCT}_{\nu_{r}}(t_{r};\theta
  \sqrt{n_{r}^{*}}) f(\theta \given t_{o}) \, \text{d} \theta}
\end{align*}
and using numerical integration as the integrals cannot be evaluated
analytically.

\subsection{Log odds ratio}
In the case of binary data, we have two independent binomial samples
\begin{align*}
  &X_{1i} \given \pi_1 \sim \Bin(n_{1i}, \pi_1)&
  &X_{2i} \given \pi_2 \sim \Bin(n_{2i}, \pi_2)&
\end{align*}
for each study $i \in \{o, r\}$, and the effect of the treatment in group 1
relative to the treatment in group 2 is quantified with the logOR
$$\theta = \log \frac{\pi_1/(1 - \pi_1)}{\pi_2/(1 - \pi_2)}.$$
To obtain a marginal likelihood that only depends on $\theta$, we need to
specify a prior for either $\pi_2$ or $\pi_1$ and marginalise over it. A
principled choice is the translation invariant Jeffreys prior,
$\pi_1, \pi_2 \sim \Be(1/2, 1/2)$. The exact marginal likelihood for the data
from study $i$ is then given by
\begin{align}
  \label{eq:binexact}
  f(x_{1i}, x_{2i} \given \theta)
  &= \int_0^1 \Bin\left(x_{1i}; n_{1i}, \left\{1 + \exp\left[-\theta - \log
    \frac{\pi_2}{1 - \pi_2}\right]\right\}^{-1}\right) \Bin(x_{2i}; n_{2i}, \pi_2)
  \nonumber
  \\ &\phantom{= \int_0^1}  \times \Be(\pi_2; 1/2, 1/2) \, \text{d} \pi_2
\end{align}
where $\Bin(x; n, \pi)$ denotes the probability mass function of the binomial
distribution with $n$ trials and probability $\pi$ evaluated at $x$, and
likewise $\Be(y; a, b)$ denotes the density function of the beta distribution
with parameters $a$ and $b$ evaluated at $y$.

There is no conjugate prior for the logOR under model~\eqref{eq:binexact}, but a
pragmatic choice is to specify a zero-mean normal prior
$\theta \given \HS \sim \Nor(0, \tau^{2})$ for the sceptic, to match with the
normal approximation as the sample size increases. For the advocacy prior, we
need to know the posterior distribution of the logOR $\theta$ based on the
original study.
Using a result from \citet{Marshall1988} combined with a change-of-variables,
the exact posterior density of the logOR $\theta$ given the original data and
Jeffreys priors on $\pi_1$ and $\pi_2$ is
\begin{align*}
  f(\theta \given x_{1o}, x_{2o}) &=
  \begin{cases}
    % C \, \exp\{(x_{1o} + 1/2) \theta\} \,
    % F(n_{1o} + 1, x_{1o} + x_{2o} + 1, n_{1o} + n_{2o} + 2, 1 - \exp(\theta))
    % & \text{for} ~ \theta < 0 \\
    % C \, \exp\{-(n_{1o} - x_{1o} + 1/2) \theta\} \,
    % F(n_{1o} + 1, n_{10} - x_{1o} + n_{2o} - x_{2o} + 1, n_{1o} + n_{2o} + 2,
    % 1 - \exp(-\theta)) & \text{for} ~ \theta > 0 \\
    C \, \exp\{e \theta\} \,
    F\left(e + f, e + g, e + f + g + h, 1 - \exp\{\theta\}\right)
    & \text{for} ~ \theta < 0 \\
    C \, \exp\{-f \theta\} \,
    F\left(e + f, f + h, e + f + g + h, 1 - \exp\{-\theta\}\right)
    & \text{for} ~ \theta > 0 \\
  \end{cases}
\end{align*}
where $F(\cdot)$ is the hypergeometric function, $e = x_{1o} + 1/2$,
$f = n_{1o} - x_{1o} + 1/2$, $g = x_{2o} + 1/2$, $h = n_{2o} - x_{2o} + 1/2$,
$C = \text{B}(e + g, f + h)/\{\text{B}(e, f) \text{B}(g, h)\}$, and
$\text{B}(\cdot, \cdot)$ is the Beta function.

Combining the previous results, we obtain
\begin{align*}
  \BFcol{0}{\text{S}}(x_{1o}, x_{2o};\tau^{2})
  &= \frac{ f(x_{1o}, x_{2o} \given 0)}{\int
   f(x_{1o}, x_{2o} \given \theta)
  \Nor(\theta; 0, \tau^{2}) \, \text{d}\theta} \\
  \BFcol{\text{S}}{\text{A}}(x_{1r}, x_{2r};\tau^{2})
  &= \frac{\int f(x_{1r}, x_{2r} \given \theta)
    \Nor(\theta; 0, \tau^{2}) \, \text{d}\theta}{
    \int f(x_{1r}, x_{2r} \given \theta)
    f(\theta \given x_{1o}, x_{2o}) \, \text{d} \theta}
\end{align*}
as an exact replacement for the Bayes factors in Box~\hyperref[box:nutshell]{1}.
Again, there are no closed form expressions for the integrals, but numerical
integration needs to be used.

% Results
% ======================================================================
\section{Application}
\label{sec:applications}
The following section will illustrate application of the sceptical Bayes factor
using data from the \emph{Social Sciences Replication Project}
\citep{Camerer2018}, provided in Table \ref{tab:ssrp}. Effect estimates were
reported on the correlation scale ($r$), which is why we applied the Fisher
$z$-transformation $\hat{\theta} = \text{tanh}^{-1}(r)$. This leads to the
transformed estimates having approximate variance
$\Var(\hat{\theta}) = 1/(n - 3)$ \citep{Fisher1921}, so the relative variance
$c$ is roughly the ratio of the replication to the original study sample size
$c \approx n_{r}/n_{o}$.
\begin{table}[!htb]
\centering
\caption{Results for data from \emph{Social Sciences Replication Project}
  \citep{Camerer2018}. Shown are relative variances
  $c = \sigma^{2}_{o}/\sigma^{2}_{r}$, relative effect estimates
  $d = \that_{r}/\that_{o}$ (computed on Fisher $z$-scale), $Q$-statistic
  $Q = (\that_{o} - \that_{r})^{2}/(\sigma^{2}_{o} + \sigma^{2}_{r})$, minimum
  Bayes factors of original and replication effect estimate ($\minBF$),
  recalibrated sceptical $p$-value ($\tilde{p}_{\text{S}}$), sceptical Bayes
  factors ($\BFs$) and replication Bayes factors ($\BFr$), the latter two
  computed using either a normal approximation or the exact likelihood of the
  data. }
\resizebox{1\textwidth}{!}{%

% latex table generated in R 4.1.1 by xtable 1.8-4 package
% Mon Aug 23 09:08:51 2021
\begin{tabular}{lrrrlllllll}
  \toprule
Original study & $c$ & $d$ & $Q$ & $\minBF_o$ & $\minBF_r$ & $\tilde{p}_{\text{S}}$ & $\text{BF}_{\text{S}}$ & $\text{BF}_{\text{S}}$ (exact) & $\text{BF}_{\text{R}}$ & $\text{BF}_{\text{R}}$ (exact) \\ 
  \midrule
Hauser et al. (2014) & 0.51 & 1.04 & 0.03 & < 1/1000 & < 1/1000 & < 0.0001 & < 1/1000 & < 1/1000 & < 1/1000 & < 1/1000 \\ 
  Aviezer et al. (2012) & 0.92 & 0.60 & 3.49 & < 1/1000 & 1/347 & < 0.0001 & 1/78 & 1/15 & 1/284 & 1/76 \\ 
  Wilson et al. (2014) & 1.33 & 0.83 & 0.28 & < 1/1000 & 1/659 & 0.0001 & 1/45 & 1/34 & < 1/1000 & < 1/1000 \\ 
  Derex et al. (2013) & 1.29 & 0.65 & 1.14 & 1/520 & 1/17 & 0.002 & 1/8.5 &  & 1/31 &  \\ 
  Gneezy et al. (2014) & 2.31 & 0.81 & 0.22 & 1/18 & 1/157 & 0.004 & 1/6.9 & 1/7.5 & 1/474 & 1/551 \\ 
  Karpicke and Blunt (2011) & 1.24 & 0.58 & 1.75 & < 1/1000 & 1/9.6 & 0.002 & 1/5.6 & 1/4.9 & 1/12 & 1/11 \\ 
  Morewedge et al. (2010) & 2.97 & 0.76 & 0.30 & 1/7.3 & 1/65 & 0.011 & 1/3.9 & 1/3.9 & 1/160 & 1/148 \\ 
  Kovacs et al. (2010) & 4.38 & 1.38 & 0.59 & 1/3.2 & < 1/1000 & 0.009 & 1/3.2 & 1/3.8 & < 1/1000 & < 1/1000 \\ 
  Duncan et al. (2012) & 7.42 & 0.57 & 1.29 & 1/12 & < 1/1000 & 0.011 & 1/3.1 & 1/3.1 & < 1/1000 & < 1/1000 \\ 
  Nishi et al. (2015) & 2.42 & 0.57 & 1.05 & 1/12 & 1/6.1 & 0.016 & 1/2.5 & 1/2.1 & 1/8.2 & 1/7 \\ 
  Janssen et al. (2010) & 0.65 & 0.48 & 3.51 & < 1/1000 & 1/3.3 & 0.003 & 1/1.6 &  & 1/1.6 &  \\ 
  Balafoutas and Sutter (2012) & 3.48 & 0.52 & 1.02 & 1/4.2 & 1/3.6 & 0.04 & 1/1.6 & 1/1.6 & 1/3.9 & 1/3.9 \\ 
  Pyc and Rawson (2010) & 9.18 & 0.38 & 1.79 & 1/3.5 & 1/7.3 & 0.061 & 1/1.2 & 1/1.2 & 1/4 & 1/3.9 \\ 
  Rand et al. (2012) & 6.27 & 0.18 & 3.96 & 1/7.1 & 1 & 0.13 &  &  & 9.6 & 9.7 \\ 
  Ackerman et al. (2010) & 11.69 & 0.23 & 2.15 & 1/2.2 & 1/1.3 & 0.15 &  &  & 3.2 & 3.3 \\ 
  Sparrow et al. (2011) & 3.50 & 0.13 & 5.80 & 1/26 & 1 & 0.19 &  &  & 29 & 30 \\ 
  Shah et al. (2012) & 11.62 & -0.05 & 4.08 & 1/2.2 & 1 & 0.66 &  &  & 25 & 26 \\ 
  Kidd and Castano (2013) & 8.57 & -0.10 & 6.83 & 1/5.7 & 1 & 0.77 &  &  & 72 & 66 \\ 
  Gervais and Norenzayan (2012) & 9.78 & -0.12 & 5.44 & 1/3 & 1 & 0.78 &  &  & 36 & 36 \\ 
  Lee and Schwarz (2010) & 7.65 & -0.11 & 6.80 & 1/5.4 & 1 & 0.79 &  &  & 65 & 64 \\ 
  Ramirez and Beilock (2011) & 4.47 & -0.09 & 19.29 & < 1/1000 & 1 & 0.85 &  &  & > 1000 & > 1000 \\ 
   \bottomrule
\end{tabular}

}
\label{tab:ssrp}
\end{table}

% % EJ Wagenmakers' analysis
% % - Ackerman et al. (2010): BFR 3.23  -- our BFR 3.2
% % -* Aviezer et al. (2012): BFR 1/57 -- our BFR 1/284
% % - Balafoutas and Sutter (2012): 1/4.27 -- our BFR 1/3.9
% % -** Derex et al. (2013): 1/3'701 -- our BFR 1/31
% % - Duncan et al. (2012): 1/2513 -- our BFR 1/3079.4
% % - Gervais & Norenzayan (2012): 36.5 -- our BFR 36
% % - Gneezy et al. (2014): 1/485.27 -- our BFR 1/474
% % -* Hauser et al. (2014): 1/10'211 -- our BFR 1/620'059
% % -** Janssen et al. (2010): 2'134'835 -- our BFR 1/1.6
% % - Karpicke & Blunt (2011): 1/11.82 -- our BFR 12
% % - Kidd & Castano (2013): 71.43 -- our BFR 72
% % - Kovacs et al. (2010): 1/132'389'305 -- our BFR 1/333'616'129
% % - Lee & Schwarz (2010): 68 -- our BFR 65
% % - Morewedge et al. (2010): 1/158 -- our BFR 1/160
% % - Nishi et al. (2015): 1/7.77 -- our BFR 1/8.2
% % - Pyc and Rawson (2010): 1/4.04 -- our BFR 1/4
% % -* Ramirez & Beilock (2011): 12'240 -- our BFR 24'996
% % - Rand et al. (2012): 9.71 -- our BFR 9.6
% % - Shah et al. (2012): 26.05 -- our BFR 25
% % - Sparrow et al. (2011): 31.65 -- our BFR 29
% % -* Wilson et al. (2014): 1/1871 -- our BFR 1/2533

For all studies except \citet{Janssen2010} and \citet{Derex2013}, the exact
approach for either SMD or logOR effect sizes from Section~\ref{sec:tdist} is
applicable. In the studies with binary data
% \citep{Balafoutas2012, Gneezy2014, Hauser2014},
computing the exact posterior using the hypergeometric function led to numerical
issues in some cases and numerical integration was used then.
% Balafoutas and Sutter - integration and hypergeo work
% Gneezy et al. - only integration works
% Hauser et al. - only hypergeo works
In most cases, the normal approximation of the likelihood seems to lead to
similar numerical results for both $\BFs$ and $\BFr$ as compared to their
counterparts based on exact likelihoods. Qualitative conclusions are the same
under both approaches and we will therefore focus on the normal approximation
due to better comparability with the remaining measures of replication success
as all of them were computed based on approximate normal likelihoods.

For the study pairs where the sceptical Bayes factor suggests a large degree of
replication success, all other methods suggest the same in every case. However,
there are also cases where there appear to be discrepancies among the methods.
For instance, the two-trials rule and the replication Bayes factor may indicate
a larger degree of replication success compared to the sceptical $p$-value and
sceptical Bayes factor. This happens for replications that show a substantial
increase in sample size but also a much smaller effect estimate compared to the
original study. For example, in \citet{Balafoutas2012} the sample size was about
$c = 3.48$ times larger in the replication, whereas the effect estimate was only $d =
0.52$ the size of the original one. The replication is
successful at $\gamma = 1/3$ with the two-trials rule
($\minBF_{o} = 1/4.2$ and $\minBF_{r} =
1/3.6$) and the replication Bayes factor
($\BFr = 1/3.9$), but not with the sceptical Bayes factor ($\BFr
= 1/1.6$) or the sceptical $p$-value
($\ps = 0.04 > 1 - \Phi(z_\gamma =
2.18) = 0.01$).

%% ps does but BFs doesn't flag RS
Discrepancies between the sceptical $p$-value and the sceptical Bayes factor
happen in situations where the replication shows an effect estimate that, although
incompatible with the sceptical prior, is also incompatible with the advocacy prior.
For example in the \citet{Janssen2010} replication, both effect estimates are
substantially larger than zero ($\that_o = 0.74$ with
$\minBF_{o} < 1/1000$ and $\that_r =
0.36$ with
$\minBF_{r} = 1/3.3$), yet the $Q$-statistic indicates some
incompatibility ($Q = 3.51$), which explains
why $\ps = 0.003$, but $\BFs =
1/1.6$ only.

Discrepancies between the replication Bayes factor and the sceptical Bayes
factor arise when the replication finding provides overwhelming evidence
against the null, whereas the original finding was less compelling.
The replication of \citet{Kovacs2010} illustrates this situation.
The original study provided only moderate evidence against the null
($\that_o = 0.49$ and
$\minBF_o
= 1/3.2$),
whereas the replication finding was more compelling
($\that_r = 0.67$ and
$\minBF_r < 1/1000$).
By construction the sceptical Bayes factor can only be as small as the minimum
Bayes factor from the original study $\minBF_o$,
which is actually attained in this case ($\BFs = 1/3.2$).
The replication Bayes factor, on the other hand, is not limited by the moderate
level of evidence from the original study and indicates decisive evidence for
the advocate ($\BFr < 1/1000$).
This illustrates that in
order to achieve a reasonable degree of replication success, the sceptical
Bayes factor requires the original study to be convincing, whereas the
replication Bayes factor only requires a compelling replication result.

% Discussion
% ======================================================================
\section{Discussion}
\label{sec:discussion}
We proposed a novel method for the statistical assessment of replicability
combining reverse-Bayes analysis with Bayesian hypothesis testing. Compared to
other methods, the sceptical Bayes factor poses more stringent requirements but
also allows for stronger statements about replication success. It ensures that
both studies provide sufficient evidence against a null effect, while also
penalising incompatibility of their effect estimates. If the replication sample
size is not too small, the sceptical Bayes factor comes with appropriate
frequentist error rates, which is often a requirement from research funders and
regulators. Asymptotic analysis of the method showed that it is information
consistent in the sense that if the sample size in both studies increases, the
sceptical Bayes factor will indicate overwhelming replication success when the
underlying effect size of the replication is not much smaller than the
underlying effect size of the original study. Finally, the sceptical Bayes
factor is the only method in our comparison which does not suffer from any form
of the shrinkage paradox, \ie replication success can never be achieved with
arbitrarily small replication effect estimates, not even when the replication
sample size becomes very large or the evidence from the original study
overwhelming.

In extreme scenarios the sceptical Bayes factor can suffer from the replication
paradox, which means that it may flag success when the replication estimate goes
in opposite direction of the original one. However, the paradox can be avoided
by truncating the advocacy prior to the direction of the original estimate. It
may also happen that the result of the replication is so inconclusive that
replication success cannot be established at any level, so the sceptical Bayes
factor does not exist. Other methods, such as the sceptical $p$-value or the
replication Bayes factor, can be used in this situation.

%% extensions and limitations
The proposed method could be extended in many ways. First, in many cases not
just one but several replication studies are conducted for one original study
\citep[\eg as in][]{Klein2014}. The Bayesian framework allows to easily extend
the sceptical Bayes factor to the ``many-to-one'' replication setting as the
likelihoods are also straightforward to compute for a sample of replication
effect estimates. Second, a multivariate generalisation would allow for effects
in the form of a vector with approximate multivariate normal likelihood which is
then combined with a sceptical $g$-prior \citep{Liang2008}. The normal prior
could also be replaced with other distributions, for example the (multivariate)
Cauchy distribution which is often the preferred prior choice for default Bayes
factor hypothesis tests \citep{Jeffreys1961}. The $g$ parameter of the $g$-prior
or the scale parameter of the Cauchy prior would then take over the role of the
relative sceptical prior variance. Third, based on the replication result one
could also compute a posterior distribution for the effect size based on a
model-average of the advocacy prior and the sceptical prior (using the variance
at the sceptical Bayes factor). This distribution would provide a formal
compromise between scepticism and advocacy of the original finding. Fourth,
while Bayes factors are an important part in Bayesian hypothesis testing, they
do not take into account the prior probabilities of the hypotheses under
consideration. It would be interesting to investigate whether the reverse-Bayes
approach could be used in a framework where priors are assigned jointly to the
hypothesis and parameter space \citep{Dellaportas2012}. Finally, an
important aspect is the design of new replication studies. An appropriate sample
size is of particular importance for a replication to be informative. We will
report in the future on sample size planning based on the sceptical Bayes
factor.

% conclusion
For a thorough assessment of replication attempts, no single metric seems to be
able to answer all important questions completely. Instead, we recommend that
researchers conduct a comprehensive statistical evaluation of replication
success.
% We advocate the reverse-Bayes
% approach as a key part of such analyses, as it
% naturally fits the replication setting, combines different notions of
% replication success, avoids various paradoxes from which other methods
% suffer, and can thereby lead to more sensible inferences and decisions.
Reverse-Bayes methods naturally fit to the replication setting, they avoid
various paradoxes from which other methods suffers, and they combine different
notions of replicability. The reverse-Bayes approach therefore leads to sensible
inferences and decisions, which is why we advocate it as a key part in the
assessment of replication success.

\section*{Software and data}
All analyses were performed in the R programming language version
4.1.1 \citep{R}. The code to
reproduce this manuscript is available at
\url{https://gitlab.uzh.ch/samuel.pawel/BFScode}. We used the implementation of
the Lambert $W$ function from the package \texttt{lamW} \citep{Adler2015},
graphics were created with the \texttt{ggplot2} package \citep{Wickham2016}, the
sceptical $p$-value and related calculations were conducted using the package
\texttt{ReplicationSuccess} available on the Comprehensive R Archive Network
% \url{https://cran.r-project.org/web/packages/ReplicationSuccess/index.html}
\citep{Held2020}. All methods are implemented in the R package \texttt{BayesRep}
which is available at \url{https://gitlab.uzh.ch/samuel.pawel/BayesRep}.

Data on effect estimates from the \emph{Social Sciences Replication Project}
\citep{Camerer2018} were downloaded from \url{https://osf.io/abu7k/},
respectively, taken from \url{https://osf.io/nsxgj/} for exact calculations.

\section*{Acknowledgements}
We thank the anonymous referees for the helpful comments and suggestions that
have considerably improved the paper. We also thank Guido Consonni, Luca La
Rocca, Malgorzata Roos, Georgia Salanti, Charlotte Micheloud, and Maria
Bekker-Nielsen Dunbar for helpful discussion and comments on drafts of the
manuscript.

\section*{Funding}
This work was supported by the Swiss National Science Foundation (project number
189295, \url{http://p3.snf.ch/Project-189295}). The funder had no role in study
design, data collection and analysis, decision to publish, or preparation of the
manuscript.

% Bibliography
% ======================================================================
\bibliographystyle{apalikedoiurl}
% \singlespacing
\bibliography{bibliography}

% Appendix
% ======================================================================
\begin{appendices}
\label{sec:appendices}

\section{Sufficiently sceptical relative prior variance}
\label{appendix:ssrv}
The sufficiently sceptical relative prior variance at level $\gamma$ is the
value $\gy \in [0, \gminBFo]$ that fulfils the condition
\begin{align}
  \label{sspv}
  \BFcol{0}{\text{S}}(\hat{\theta}_o; \gy)
  &= \gamma.
\end{align}
Substituting~\eqref{sspv} and rearranging terms, we obtain
\begin{align}
  \label{sspv2}
  \sqrt{1 + \gy} \cdot
  \exp\left\{-\frac{1}{2} \cdot
  \frac{\gy}{1 + \gy} \cdot z^2_o\right\}
  &= \gamma \nonumber \\
  \iff \frac{1}{\gamma} \cdot \exp\left\{-\frac{z_o^2}{2}\right\}
  &= \frac{1}{\sqrt{1 + \gy}} \exp\left\{-\frac{1}{2} \cdot
  \frac{z_o^2}{1 + \gy}\right\}. \nonumber
  \\
  \intertext{Squaring both sides and multiplying by $-z_o^2$, this becomes}
  \iff -\frac{z_o^2}{\gamma^2} \cdot \exp\left\{-z_o^2\right\}
  &= -\frac{z_o^2}{1 + \gy}
  \exp\left\{-\frac{z_o^2}{1 + \gy}\right\}.
\end{align}
This is a transcendental equation that cannot be explicitly solved in terms of
elementary functions. However, if we set $q = -z_o^2/(1 + \gy)$
then~\eqref{sspv2} becomes
\begin{align*}
  -\frac{z_o^2}{\gamma^2} \cdot \exp\left\{-z_o^2\right\}
  &= q \cdot \exp\left\{q\right\}.
\end{align*}
The solution for $q$ (and consequently for $\gy$) can be explicitly computed
with
\begin{align}
\label{eq:solutionvss}
  q &= \lw{-1}\left(-\frac{z_o^2}{\gamma^2} \cdot \exp\left\{-z_o^2\right\}\right)
  \nonumber \\
  \gy &=
  \begin{cases}
    -\dfrac{z_o^2}{q} - 1 & ~~ \text{if} ~ -\dfrac{z_o^2}{q} \geq 1 \\
    \text{undefined} & ~~ \text{else}
  \end{cases}
\end{align}
where $\lw{-1}(\cdot)$ is the branch of the Lambert $W$ function that satisfies
$W(y) \leq -1$ for $y \in [-e^{-1}, 0)$, ensuring that $\gy \leq \gminBFo$. See
Appendix~\ref{appendix:lambertW} for details about the Lambert $W$ function. For
some $z_o$, equation \eqref{sspv2} can also be satisfied for negative $\gy$,
which is why we need to add the condition $-z_o^2/q \geq 1$ in equation
\eqref{eq:solutionvss}, such that $\gy$ is a valid relative variance.

As $z_{o}^{2}$ becomes larger, the argument to the Lambert function
$x = -z_{o}^{2}\exp(-z_{o}^{2})/\gamma^{2}$ will approach zero, so the the
approximation $\lw{-1}(x) \approx \log(-x) - \log(-\log(-x))$ can be applied
\citep[p. 350]{Corless1996}. This leads to
\begin{align*}
  \gy \approx \frac{z_{o}^{2}}{z_{o}^{2} + \log \gamma^{2} - \log z_{o}^{2} +
  \log\{z_{o}^{2} + \log \gamma^{2} - \log z_{o}^{2}\}} - 1.
\end{align*}
We can see that $\gy \downarrow 0$ when $\gamma$ remains fixed and
$z_{o}^{2} \to \infty$, which means that the sufficiently sceptical relative
prior variance converges to zero for increasingly compelling evidence from the
original study.

\section{The Lambert $W$ function}
\label{appendix:lambertW}
\begin{figure}[!htb]
\begin{knitrout}
\definecolor{shadecolor}{rgb}{0.969, 0.969, 0.969}\color{fgcolor}
\includegraphics[width=\maxwidth]{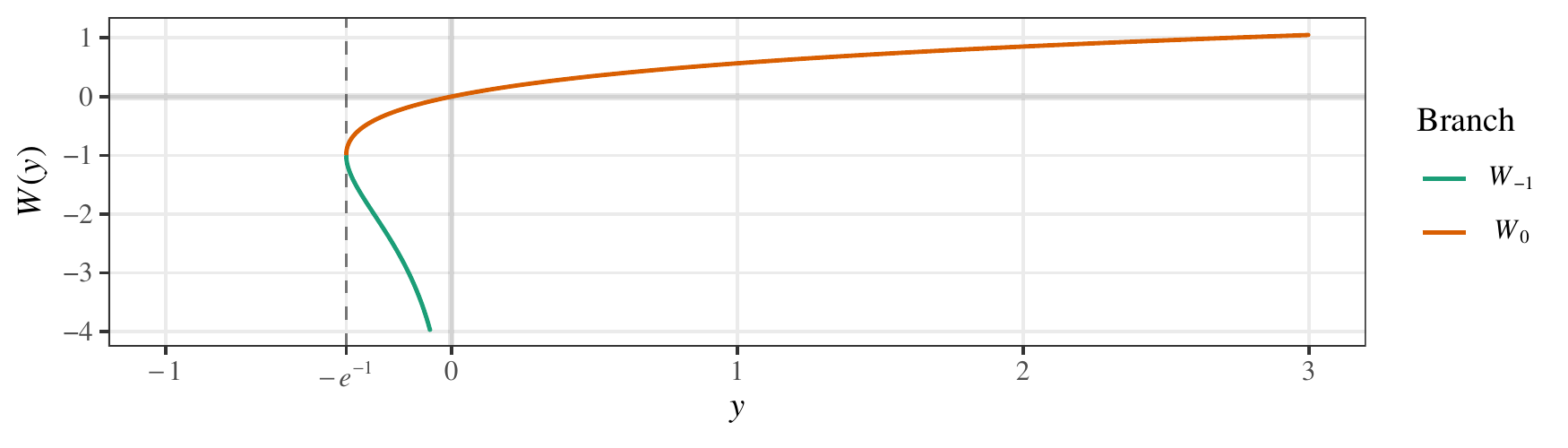}
\end{knitrout}
\caption{Lambert $W$ function for real argument $y$.}
\label{fig:lambertW}
\end{figure}

The Lambert $W$ function \citep{Corless1996} is defined as the function
$W(\cdot)$ satisfying
$$W(y) \cdot \exp\{W(y)\} = y$$
and it is also known as ``product logarithm`` since it returns the number which
plugged in the exponential function and then multiplied by itself produces $y$.
For real $y$, $W(y)$ is only defined for $y \geq - e^{-1}$ and for
$y \in [-e^{-1}, 0)$ the function has two branches that are commonly denoted by
$\lw{0}(\cdot)$, the branch with $W(y) \geq -1$, and $\lw{-1}(\cdot)$, the
branch with $W(y) \leq -1$ (see Figure \ref{fig:lambertW} for an illustration).

\section{Computation of the sceptical Bayes factor}
\label{appendix:bfs}
From the definition of the sceptical Bayes factor it is apparent that $\BFs$ is
either
\begin{enumerate}
  \item \label{undef} undefined,
  if
  $\BFcol{\text{S}}{\text{A}}(\hat{\theta}_r; g) >
  \BFcol{\text{0}}{\text{S}}(\hat{\theta}_o; g)$ for all $g \in [0,
    \gminBFo]$

  \item \label{lowbound} $\BFs = \minBF_{o}$, if
  $\BFcol{\text{S}}{\text{A}}(\hat{\theta}_r; \gminBFo) <
  \BFcol{0}{\text{S}}(\hat{\theta}_o; \gminBFo)$

  \item \label{intersection}
  $\BFs = \inf_{\gy}
  \left\{\gamma : \BFcol{\text{S}}{\text{A}}(\hat{\theta}_r; \gy)
  = \gamma\right\}$,
  the height of the lowest intersection of
  $\BFcol{0}{\text{S}}(\hat{\theta}_o; \gy) = \gamma$ and
  $\BFcol{\text{S}}{\text{A}}(\hat{\theta}_r; \gy)$
  in $\gy$ otherwise
\end{enumerate}
Whether $\BFs$ attains the lower bound $\minBF_{o}$ (condition \ref{lowbound})
can be checked by evaluating if
$\BFcol{\text{S}}{\text{A}}(\hat{\theta}_r; \gminBFo) < \minBF_{o}$ and setting
$\BFs = \minBF_{o}$ if it is the case. For condition \ref{intersection}, we know
that the intersections satisfy
\begin{align*}
  \BFcol{\text{S}}{\text{A}}(\hat{\theta}_r; g_*) &=
  \BFcol{0}{\text{S}}(\hat{\theta}_o; g_*) \\
  \sqrt{\frac{1/c + 1}{1/c + g_*}} \cdot
  \exp\left\{-\frac{z_{o}^{2}}{2}\left(\frac{d^2}{1/c + g_*} -
  \frac{\left(1 - d\right)^2}{1/c + 1} \right)\right\}
  &= \sqrt{1 + g_*} \cdot
  \exp\left\{-\frac{1}{2} \cdot \frac{g_*}{1 + g_*} \cdot z^2_o\right\}
\end{align*}
which is equivalent to
\begin{align}
  \label{eq:bfsroot}
  \sqrt{\frac{1}{(1 + g_*)(1/c + g_*)}}\cdot\exp\left\{-\frac{z_{o}^{2}}{2}\left(
  \frac{1}{1 + g_*} + \frac{d^{2}}{1/c + g_*}\right)\right\}
  &= \sqrt{\frac{1}{1/c + 1}}\cdot\exp\left\{-\frac{z_{o}^{2}}{2}\left(
  1 + \frac{(1 - d)^{2}}{1/c + 1}\right)\right\}.
\end{align}
This is a transcendental equation that has no closed-form solution for $g_*$ in
terms of elementary functions, but root-finding algorithms can be used to
compute it. However, when $c = 1$, equation~\eqref{eq:bfsroot} simplifies
\begin{align}
  \label{eq:bfsroot2}
  \frac{1}{1 + g_*}\cdot\exp\left\{-\frac{z_{o}^{2}}{2} \cdot
  \frac{1 + d^{2}}{1 + g_*}\right\}
  &= \frac{1}{\sqrt{2}}\cdot\exp\left\{-\frac{z_{o}^{2}}{2}\left(
  1 + \frac{(1 - d)^{2}}{2}\right)\right\}.
\end{align}
Multiplying~\eqref{eq:bfsroot2} by $-z_{o}^{2}(1 + d^{2})/2$ and applying the
Lambert $W$ function leads to
\begin{align}
  \label{eq:gintersect}
  % q &= W\left(-\frac{z_A^2}{\sqrt{2}} \cdot
  % \exp\left\{-\frac{z_o^2}{2} \left(1 + \frac{(1 - d)^2}{2}\right)\right\}\right) \\
  % g_* &=
  % \begin{cases}
  %   \dfrac{-z_A^2}{q} - 1 & ~~ \text{if} ~ \dfrac{-z_A^2}{q} \geq 1 \\
  %   \text{undefined} & ~~ \text{else},
  % \end{cases}
  k &= W\left(-\frac{z_{o}^{2}}{\sqrt{2}} \cdot \frac{1 + d^{2}}{2} \cdot
      \exp\left\{-\frac{z_o^2}{2} \left[1 + \frac{(1 - d)^2}{2}\right]\right\}\right)
  \nonumber \\
  g_* &=
  \begin{cases}
    -\dfrac{z_o^2}{k}\cdot \dfrac{1 + d^{2}}{2} - 1 & ~~ \text{if} ~
    -\dfrac{z_o^2}{k}\cdot\dfrac{1 + d^{2}}{2} \geq 1 \\
    \text{undefined} & ~~ \text{else},
  \end{cases}
\end{align}
with the condition that $-z_o^2(1 + d^{2})/(2k) \geq 1$ such that $g_*$ is a
valid relative variance, as the equation may otherwise be satisfied for negative
$g_*$. Since the argument to $W(\cdot)$ is real and negative (if $z_{o}\neq 0$),
the branches $\lw{-1}(\cdot)$ and $\lw{0}(\cdot)$ both provide solutions that
can fulfil the equation (assuming the argument is not smaller than $-e^{-1}$
which would mean that there are no intersections). It must also hold that
$g_{*} \leq g_{\minBF_{o}} = \max\{z_{o}^{2} - 1, 0\}$ for $g_{*}$ to be a valid
sufficiently sceptical prior variance. Hence, when $|d| \leq 1$, the $g_{*}$
from~\eqref{eq:gintersect} can only be computed with the $\lw{-1}(\cdot)$
branch, whereas for $|d| > 1$ and when $-k \geq (1 + d^{2})/2$ the solution
$g_{*}$ is computed from the $\lw{0}(\cdot)$ branch. Plugging the relative prior
variance $g_{*}$ from~\eqref{eq:gintersect} into the Bayes factor
from~\eqref{eq:BFo}, we obtain the expression for the sceptical Bayes factor
in~\eqref{eq:BFsc1}.

\section{Bayes factor with truncated advocacy prior}
\label{app:BFtrunc}
For now assume $\that_o > 0$. The marginal likelihood of the replication effect
estimate $\that_r \given \theta \sim \Nor(\theta, \sigma^2_r)$ under the
truncated advocacy prior
$\HAp \colon \theta \sim \Nor(\that_o, \sigma^2_o) \mathbb{1}_{(0, \infty)}(\theta)$
is
\begin{align}
  f(\that_r \given \HAp)
  &= \int_{-\infty}^{+\infty} f(\that_r \given \theta) f(\theta \given
  \HAp) \, \text{d}\theta \nonumber \\
  &= \int_{-\infty}^{+\infty}
  \frac{\mathbb{1}_{(0, \infty)}(\theta)}{2\pi\sigma_r\sigma_o\Phi(z_o)}
  \exp\left\{-\frac{1}{2}\left[\frac{(\that_o - \theta)^2}{\sigma^2_r}
  + \frac{(\theta - \that_o)^2}{\sigma^2_o}\right]\right\}
  \, \text{d}\theta \nonumber \\
  &=
  \frac{1}{2\pi\sigma_r\sigma_o\Phi(z_o)}
  \exp\left\{-\frac{1}{2}\frac{(\that_r - \that_o)^2}{\sigma^2_r + \sigma^2_o}\right\}
   \underbrace{\int_{0}^{+\infty}
   \exp\left\{-\frac{1}{2}\frac{(\theta - \frac{\that_o/\sigma^2_o +
       \that_r/\sigma^2_r}{1/\sigma^2_o + 1/\sigma^2_r})^2}{\left(
     1/\sigma^2_o + 1/\sigma^2_r\right)^{-1}}\right\}
   \, \text{d}\theta}_{=\sqrt{\frac{2\pi}{1/\sigma^2_o + 1/\sigma^2_r}}
   % \Phi\left(\frac{z_o + z_r\sqrt{c}}{\sqrt{1 + c}}\right)} \nonumber \\
  \Phi\left(\frac{z_o(1 + dc)}{\sqrt{1 + c}}\right)} \nonumber \\
   &= \frac{1}{\sqrt{2\pi(\sigma_r^2 + \sigma_o^2)}}
   \exp\left\{-\frac{1}{2}\frac{(\that_r - \that_o)^2}{\sigma^2_r + \sigma^2_o}
     % \right\}  \frac{\Phi\left(\frac{z_o + z_r\sqrt{c}}{\sqrt{1 + c}}\right)}{
     \right\}  \frac{\Phi\left(\frac{z_o(1 + dc)}{\sqrt{1 + c}}\right)}{
     \Phi(z_o)}.
   \label{eq:truncmarg}
\end{align}
With a similar argument one can show that this result holds for any $\that_o$ if
the last factor in \eqref{eq:truncmarg} is changed to
\begin{align*}
  % \frac{\Phi\left\{\text{sign}(z_o) \frac{z_o + z_r\sqrt{c}}{\sqrt{1 + c}}
  \frac{\Phi\left\{\text{sign}(z_o) \frac{z_o(1 + dc)}{\sqrt{1 + c}}
  \right\}}{\Phi\left\{\abs{z_o}\right\}}.
\end{align*}
By dividing the marginal likelihood of the replication data under the
sceptical prior by the marginal likelihood under the truncated advocacy prior,
the Bayes factor in \eqref{eq:BFrtrunc} is obtained.

\section{The shrinkage paradox}
\label{app:shrinkage}
We want to investigate what happens to the replication success regions as the
relative variance $c$ and the squared original $z$-value $z_{o}^{2}$ (a monotone
transformation of the original Bayes factor $\minBF_{o}$) become larger.
Ignoring the success regions on the wrong side of zero (due to the replication
paradox), the minimum relative effect estimates $\dmin$ as shown in
Section~\ref{sec:comparison} are given by

\begin{align*}
    &\dmin^{\scriptscriptstyle \BFs} =
      \frac{1/c + \gy}{\gy - 1} +
      \sqrt{\frac{\log\left[\frac{1/c + 1}{(1/c + \gy)(1 + \gy)}\right]/z_{o}^{2}
      + \frac{\gy}{1 + \gy} + \frac{1}{1 - \gy}}{(1 - \gy)/\left[(1/c + \gy)(1/c + 1)\right]}}&
  &\text{(sceptical Bayes factor)} \\
    &\dmin^{\scriptscriptstyle \text{2TR}} =
      \frac{z_{\gamma}}{z_{o}\sqrt{c}}&
  &\text{(two-trials rule)} \\
    &\dmin^{\scriptscriptstyle \BFr} =
      \sqrt{\left[1 + \frac{\log(1 + c) - 2\log \gamma}{z_{o}^{2}}\right] \frac{1/c + 1}{c}} - \frac{1/c + 1}{1 + c} &
  &\text{(replication Bayes factor)} \\
    &\dmin^{\scriptscriptstyle \ps} =
      \sqrt{\frac{1/c + 1/(z_{o}^{2}/z_{\gamma}^{2} - 1)}{z_{o}^{2}/z_{\gamma}^{2}}}&
  &\text{(sceptical $p$-value)} \\
\end{align*}
where for the sceptical Bayes factor it was assumed that $\gy > 1$ (otherwise
the plus before the square root term needs to be replaced by a minus).

For the sceptical Bayes factor, we obtain
\begin{align*}
  &\lim_{c \to \infty} \dmin^{\scriptscriptstyle \BFs} =
  \frac{\gy}{\gy - 1} +
    \sqrt{\frac{\log\left[\frac{1}{\gy(1 + \gy)}\right]/z_{o}^{2}
    + \frac{\gy}{1 + \gy} + \frac{1}{1 - \gy}}{(1 - \gy)/\gy}}&
  &\text{and}&
  &\lim_{z_{o}^{2}\to \infty} \dmin^{\scriptscriptstyle \BFs} =
    \sqrt{\frac{1/c + 1}{c}} - \frac{1}{c}&
\end{align*}
where for the second limit we used that $\lim_{z_{o}^{2} \to \infty} \gy = 0$
for a fixed level $\gamma$ (Appendix~\ref{appendix:ssrv}). So the sceptical
Bayes factor does not suffer from any form of the shrinkage paradox. The limits
for the two-trials rule are given by
\begin{align*}
  &\lim_{c \to \infty} \dmin^{\scriptscriptstyle \text{2TR}} = 0&
  &\text{and}&
  &\lim_{z_{o}^{2}\to \infty} \dmin^{\scriptscriptstyle \text{2TR}} = 0&
\end{align*}
so the two-trials rule suffers from both forms of the shrinkage paradox. For the
sceptical $p$-value, we obtain
\begin{align*}
  &\lim_{c \to \infty} \dmin^{\scriptscriptstyle \ps} =
    \sqrt{\frac{z_{\gamma}^{2}}{z_{o}^{2}(z_{o}^{2}/z_{\gamma}^{2} - 1)}}&
  &\text{and}&
  &\lim_{z_{o}^{2}\to \infty} \dmin^{\scriptscriptstyle \ps} = 0&
\end{align*}
thus, the sceptical $p$-value suffers from the shrinkage paradox at original.
Finally, the limits for the replication Bayes factor are
\begin{align*}
  &\lim_{c \to \infty} \dmin^{\scriptscriptstyle \BFr} = 0 &
  &\text{and}&
  &\lim_{z_{o}^{2}\to \infty} \dmin^{\scriptscriptstyle \BFr} =
    \sqrt{\frac{1/c + 1}{c}} - \frac{1/c + 1}{1 + c}&
\end{align*}
which means that the replication Bayes factor suffers from the shrinkage paradox
at replication.

\section{Probability of replication success with the sceptical Bayes factor}
\label{app:powerBFs}
Conditional on the original study, the probability for replication success at
level $\gamma$ is given by the probability of \eqref{eq:RSCond-d}. This event
involves $z_r = dz_o \sqrt{c}$ as the only random quantity if the original study
has been completed. Assuming a normal distribution
\begin{align*}
  z_r \given z_o, c \sim \Nor(\mu_{z_r}, \sigma^2_{z_r})
\end{align*}
which may depend on $z_o$ and $c$ encompasses the typical scenarios under which
one would want to compute the probability for replication success. For example,
under the null hypothesis ($H_{0} \colon \theta = 0$), we have $\mu_{z_r} = 0$
and $\sigma^2_{z_r} = 1$. For conditional power we assume the underlying effect
size equals the original effect estimate ($\theta = \that_o$) and therefore
$\mu_{z_r} = z_o \sqrt{c}$ and $\sigma^2_{z_r} = 1$. Finally, predictive power
is obtained by using the predictive distribution based on the advocacy prior
($\HA \colon \theta \sim \Nor(\that_o, \sigma^2_o)$) and thus
$\mu_{z_r} = z_o \sqrt{c}$ and $\sigma^2_{z_r} = 1 + c$.

Applying some algebraic manipulations to \eqref{eq:RSCond-d}, the
probability for replication success at level $\gamma$ can be computed by
\begin{align}
  \label{eq:powerBFs}
  \Pr\left(\BFs \leq \gamma \given z_o, c\right) =
  \begin{cases}
    \Pr(\chi^2_{1,\lambda} \geq  A/[B \sigma^2_{z_r}])
    & \text{for} ~ \gy < 1 \\
    \Phi(\text{sign}(z_o) \left\{\mu_{z_r} - D\right\}/\sigma_{z_r})
    & \text{for} ~ \gy = 1 \\
    \Pr(\chi^2_{1,\lambda} \leq A/[B \sigma^2_{z_r}])
    & \text{for} ~ \gy > 1
  \end{cases}
\end{align}
with non-centrality parameter $\lambda = (\mu_{z_r} - M)^2/\sigma^2_{z_r}$
and
\begin{align*}
  A &= \log \left\{ \frac{1/c + 1}{(1/c + \gy)(1 + \gy)}\right\} +
  z_o^2 \left\{\frac{\gy}{1 + \gy} + \frac{1}{1 - \gy} \right\}, \\
  B &= \frac{1 - \gy}{(1 + c\gy)(1/c + 1)}, \\
  D &= \frac{z_o^2 \{1/2 + 1/(1/c + 1)\} - \log 2}{2 z_o \sqrt{c}} (1 + c) \\
  M &= \frac{z_o(1 + c\gy)}{\sqrt{c}(\gy - 1)}.
\end{align*}
The probability is zero, if the original $z$-value $|z_o|$ is not large enough
such that the sufficiently sceptical relative prior variance $\gy$ can be
computed for level $\gamma$ with \eqref{ggamma}.

\section{Probability of replication success with the replication Bayes factor}
\label{app:powerBFr}
The probability of $\BFr < \gamma$ is equivalent to the probability of
\begin{align}
  \label{eq:BFrsuccess}
  \log(1 + c) - z_{r}^{2} + \frac{(z_{r} - z_{o}\sqrt{c})^{2}}{1 + c}
  \leq  2 \log \gamma
\end{align}
Applying some algebraic manipulations to \eqref{eq:BFrsuccess} and assuming a
normal distribution for $z_{r}$ as in Appendix~\ref{app:powerBFs} leads to
\begin{align*}
  \Pr\left(\BFr \leq \gamma \given z_o, c\right) =
  \Pr\left(\chi^2_{1,\lambda} \geq \left\{z_{o}^{2} + \log(1 + c) - \log \gamma^{2}
  \right\} [1 + 1/c] / \sigma^2_{z_r}\right)
\end{align*}
with non-centrality parameter $\lambda = (\mu_{z_r} + z_{o}/\sqrt{c})^2/\sigma^2_{z_r}$.

\end{appendices}

\end{document}